%% file: main.tex
\documentclass[acmsmall,nonacm]{acmart}

\AtBeginDocument{%
  }

\setcopyright{none}
\usepackage{xspace}
\usepackage{listings}
\usepackage{pgf}
\usepackage{tikz}
\usetikzlibrary{arrows.meta,positioning,fit,calc,backgrounds}
\usepackage[vlined,ruled,linesnumbered]{algorithm2e}
\usepackage{caption}
\usepackage{subcaption}
\usepackage{multirow}
\usepackage{hyperref}
\usepackage{pgfplots}
\usepackage{pgfplotstable}
\usepackage{microtype}
\usepackage{balance}
\usepackage{tabularx}
\usepackage{colortbl}
\usepackage{enumitem}
\usepackage[norndcorners,customcolors,nofill]{hf-tikz}
\usepackage{xstring}
\usepackage{adjustbox}
\usepackage{booktabs}
\usepackage{amsthm}
\usepackage[leftcaption]{sidecap}
\usepackage{siunitx}
\usepackage{wrapfig}

\usepackage{solidity}
\newcommand\secref[1]{Sect.~\ref{#1}}

\newcommand\figref[1]{Fig.~\ref{#1}}

\newcommand\tabref[1]{Tab.~\ref{#1}}
\newcommand\algoref[1]{Alg.~\ref{#1}}

\newcommand\tool{\textsc{Soliduzz}\xspace}
\newcommand\circil{\textsc{CircIL}\xspace}
\newcommand\solc{\texttt{solc}\xspace}
\newcommand\solx{\texttt{solx}\xspace}
\newcommand\vyper{\texttt{vyper}\xspace}
\newcommand\erwin{\textsc{Erwin}\xspace}
\newcommand\fuzzol{\textsc{Fuzzol}\xspace}
\newcommand\solsmith{\textsc{Solsmith}\xspace}
\newcommand{\efsolsmith}{\textsc{SolSmith}\xspace}
\newcommand{\acf}{\textsc{AFL-compiler-fuzzer}\xspace}
\newcommand\csmith{\textsc{Csmith}\xspace}

\newcommand{\positive}[1]{\textcolor{evalgreen}{#1}}
\newcommand{\negative}[1]{\textcolor{evalred}{#1}}

\newcommand\code[1]{\inlinesolidity{#1}}
\newcommand{\tcode}[1]{{\ttfamily\footnotesize #1}}

\SetAlgorithmName{Alg.}{Alg.}{List of Algorithms}

\newcommand\bug[1]{%
\IfEqCase{#1}{%
{1}{#1}% XLSX #24 (dear-fowl, solx): selector miscompilation; fixed; private
{2}{#1}% XLSX #17 (mint-fox, solx): invalid opcode with >14 params, non-via-IR; fixed
{3}{\href{https://github.com/NomicFoundation/solx/issues/150}{#1}}% XLSX #21 (wired-sunfish, solx): assertion violation not reproducible with solc; fixed
{4}{#1}% XLSX #18 (stirred-martin, solx): untitled miscompilation; fixed; private
{5}{#1}% XLSX #16 (solx): EVM backward propagation stackification ICE; confirmed; private
{6}{#1}% XLSX #19 (solx): untitled crash; ICE; fixed; private
{7}{#1}% XLSX #20 (solx): untitled hang; fixed; private
{8}{\href{https://github.com/NomicFoundation/solx/issues/136}{#1}}% XLSX #22 (solx): crash on fuzzed Solidity code; fixed
{9}{\href{https://github.com/NomicFoundation/solx/issues/132}{#1}}% XLSX #23 (solx): linker crash, Invalid ELFCLASS; fixed
{10}{#1}% XLSX #15 (solc): dirty-bit handling differences; confirmed; fix pending; bug bounty
{11}{\href{https://github.com/argotorg/solidity/issues/16797}{#1}}% XLSX #14 (solc): SSA-CFG library placeholder miscompilation; confirmed
{12}{\href{https://github.com/argotorg/solidity/issues/16593}{#1}}% XLSX #9 (solc): ICE with bytecode >2^16 bytes on EOF; confirmed
{13}{\href{https://github.com/argotorg/solidity/issues/16155}{#1}}% XLSX #10 (solc): Yul expression simplifier produces invalid Yul; fixed
{14}{\href{https://github.com/argotorg/solidity/issues/16139}{#1}}% XLSX #11 (solc): stack-layout assertion after 1000 iterations; confirmed
{15}{\href{https://github.com/argotorg/solidity/issues/16715}{#1}}% XLSX #12 (solc): SSA-CFG assertion failure in StackUtils; fixed
{16}{\href{https://github.com/argotorg/solidity/issues/16712}{#1}}% XLSX #13 (solc): SSA-CFG assertion failure in StackLayoutGenerator; confirmed
{17}{\href{https://github.com/vyperlang/vyper/issues/4943}{#1}}% XLSX #1 (vyper): BOT has no lower bound; ICE; fixed
{18}{\href{https://github.com/vyperlang/vyper/issues/4930}{#1}}% XLSX #2 (vyper): AssertionError in _stack_reorder; fixed
{19}{\href{https://github.com/vyperlang/vyper/issues/4933}{#1}}% XLSX #3 (vyper): spurious TypeError with venom; fixed
{20}{\href{https://github.com/vyperlang/vyper/issues/4934}{#1}}% XLSX #4 (vyper): CompilerPanic, missing symbols; confirmed
{21}{\href{https://github.com/vyperlang/vyper/issues/4931}{#1}}% XLSX #5 (vyper): unspecific AssertionError with venom; confirmed
{22}{\href{https://github.com/vyperlang/vyper/issues/4920}{#1}}% XLSX #6 (vyper): KeyError with experimental codegen; fixed
{23}{\href{https://github.com/vyperlang/vyper/issues/4971}{#1}}% XLSX #7 (vyper): RecursionError with venom; confirmed
{24}{\href{https://github.com/vyperlang/vyper/issues/4976}{#1}}% XLSX #8 (vyper): ValueError with venom and -O 3; confirmed
{25}{\href{https://github.com/foundry-rs/foundry/issues/11615}{#1}}% XLSX #25 (forge): forge build --ast recursion limit exceeded; confirmed
{26}{\href{https://github.com/argotorg/hevm/issues/785}{#1}}% XLSX #26 (hevm): symbolic execution silently exits with code 1; fixed
{27}{\href{https://github.com/argotorg/hevm/issues/1066}{#1}}% XLSX #27 (hevm): spurious counterexample; precision; fixed
{28}{\href{https://github.com/argotorg/hevm/issues/786}{#1}}% XLSX #28 (hevm): symbolic execution index out of bounds; fixed
{29}{\href{https://github.com/argotorg/hevm/issues/1086}{#1}}% XLSX #29 (hevm): spurious counterexample; precision; open
}[\PackageError{bug}{Undefined option to bug: #1}{}]%
}

\newcommand\fix[1]{%
\IfEqCase{#1}{%
{A}{\href{https://github.com/matter-labs/era-compiler-llvm/pull/914}{#1}}% bug 1; XLSX #24 (dear-fowl, solx): fixed
{B}{\href{https://github.com/NomicFoundation/solx/pull/431}{#1}}% bug 2; XLSX #17 (mint-fox, solx): fixed
{C}{\href{https://github.com/NomicFoundation/solx/pull/173}{#1}}% bug 3; XLSX #21 (wired-sunfish, solx): fixed
{D}{\href{https://github.com/NomicFoundation/solx-llvm/pull/109}{#1}}% bug 4; XLSX #18 (stirred-martin, solx): fixed
{E}{\href{https://github.com/NomicFoundation/solx/pull/431}{#1}}% bug 6; XLSX #19 (solx crash): fixed
{F}{\href{https://github.com/NomicFoundation/solx/pull/454}{#1}}% bug 7; XLSX #20 (solx hang): fixed
{G}{\href{https://github.com/matter-labs/era-compiler-llvm/pull/894}{#1}}% bug 8; XLSX #22 (solx ICE): fixed
{H}{\href{https://github.com/NomicFoundation/solx/pull/134}{#1}}% bug 9; XLSX #23 (solx linker ICE): fixed
{I}{\href{https://github.com/argotorg/solidity/pull/16911}{#1}}% bug 10; XLSX #15 (solc dirty-bit handling): fix pending
{J}{\href{https://github.com/argotorg/solidity/pull/16161}{#1}}% bug 13; XLSX #10 (solc expression simplifier): fixed
{K}{\href{https://github.com/argotorg/solidity/pull/16817}{#1}}% bug 15; XLSX #12 (solc StackUtils assertion): fixed
{L}{\href{https://github.com/vyperlang/vyper/pull/4827}{#1}}% bug 17; XLSX #1 (vyper BOT lower bound): fixed
{M}{\href{https://github.com/vyperlang/vyper/pull/4936}{#1}}% bug 18; XLSX #2 (vyper _stack_reorder): fixed
{N}{\href{https://github.com/vyperlang/vyper/pull/4941}{#1}}% bug 19; XLSX #3 (vyper spurious TypeError): fixed
{O}{\href{https://github.com/vyperlang/vyper/pull/4921}{#1}}% bug 22; XLSX #6 (vyper experimental codegen KeyError): fixed
{P}{\href{https://github.com/argotorg/hevm/pull/797}{#1}}% bug 26; XLSX #26 (hevm silent exit): fixed
{Q}{\href{https://github.com/argotorg/hevm/pull/1071}{#1}}% bug 27; XLSX #27 (hevm spurious counterexample): fixed
{R}{\href{https://github.com/argotorg/hevm/pull/798}{#1}}% bug 28; XLSX #28 (hevm index out of bounds): fixed
{S}{\href{https://github.com/argotorg/hevm/pull/1094}{#1}}% bug 29; XLSX #29 (hevm spurious counterexample): fixed
}[\PackageError{fix}{Undefined option to fix: #1}{}]%
}

\definecolor{gray}{rgb}{0.5, 0.5, 0.5}
\definecolor{light-gray}{gray}{0.77}
\definecolor{BrickRed}{rgb}{0.8, 0.25, 0.33}
\definecolor{Black}{rgb}{0.0, 0.0, 0.0}
\definecolor{DarkBlue}{rgb}{0.0, 0.0, 0.55}
\definecolor{Crimson}{rgb}{0.86, 0.08, 0.24}
\definecolor{SlateGrey}{rgb}{0.44, 0.5, 0.56}
\definecolor{lightorange}{HTML}{FFB74D}
\definecolor{blue}{rgb}{0.0, 0.0, 1.0}
\definecolor{magenta}{rgb}{0.79, 0.08, 0.48}
\definecolor{slateblue}{RGB}{44, 62, 80}
\definecolor{lightslate}{RGB}{242, 244, 244}
\definecolor{accentblue}{RGB}{52, 152, 219}
\definecolor{teal}{rgb}{0.0, 0.5, 0.5}
\definecolor{evalgreen}{RGB}{28,105,74}
\definecolor{evalred}{RGB}{145,38,50}

\lstdefinestyle{basic}{%
  language         = Java,%
  alsoletter       = -,%
  keywordstyle     = [2]\color{teal}\bfseries,%
  morekeywords     = [3]{},%
  keywordstyle     = [3]\color{BrickRed}\bfseries,%
  keywordstyle     = \bfseries\color{DarkBlue},%
  comment          = [l]{//},%
  commentstyle     = \ttfamily\color{Black!50}\footnotesize\lst@ifdisplaystyle\scriptsize\fi,%
  basicstyle       = \ttfamily\footnotesize\lst@ifdisplaystyle\scriptsize\fi,%
  emph             = {},%
  emphstyle        = {\color{teal}\bfseries},%
  stringstyle      = \color{BrickRed},%
  columns          = [c]fixed,%
  aboveskip        = 0mm,%
  belowskip        = 2mm,%
  keepspaces       = true,%
  mathescape       = true,%
  escapechar       = ?,%
  tabsize          = 2,%
  numbers          = left,%
  numberstyle      = \tiny\color{Black!70},%
  numbersep        = 4pt,%
  stepnumber       = 1,%
  firstnumber      = 1,%
  showstringspaces = false,%
  captionpos       = b,%
  extendedchars    = true,%
  upquote          = true,%
  abovecaptionskip = 0mm,%
  belowcaptionskip = 0mm,%
  xleftmargin      = 5mm,%
  moredelim        = **[is][{\btHL[fill=light-gray]}]{°}{°},
  morecomment      = [s]{/*}{*/}
}

\lstdefinestyle{pseudo}{%
  language         = Python,%
  alsoletter       = -,%
  keywordstyle     = [2]\color{teal}\bfseries,%
  morekeywords     = [3]{},%
  keywordstyle     = [3]\color{BrickRed}\bfseries,%
  morekeywords     = [4]{@staticmethod},%
  keywordstyle     = [4]\color{Black!60}\bfseries,%
  keywordstyle     = \bfseries\color{DarkBlue},%
  commentstyle     = \ttfamily\color{Black!50}\footnotesize\lst@ifdisplaystyle\scriptsize\fi,%
  basicstyle       = \ttfamily\footnotesize\lst@ifdisplaystyle\scriptsize\fi,%
  emph             = {},%
  emphstyle        = {\color{teal}\bfseries},%
  stringstyle      = \color{BrickRed},%
  columns          = [c]fixed,%
  aboveskip        = 0mm,%
  belowskip        = 2mm,%
  keepspaces       = true,%
  mathescape       = true,%
  escapechar       = ?,%
  tabsize          = 2,%
  numbers          = left,%
  numberstyle      = \tiny\color{Black!70},%
  numbersep        = 4pt,%
  stepnumber       = 1,%
  firstnumber      = 1,%
  showstringspaces = false,%
  captionpos       = b,%
  extendedchars    = true,%
  upquote          = true,%
  abovecaptionskip = 0mm,%
  belowcaptionskip = 0mm,%
  xleftmargin      = 5mm,%
  moredelim        = **[is][{\btHL[fill=light-gray]}]{°}{°},
  morecomment      = [s]{/*}{*/}
}
\begin{document}

\title{\tool: Finding Miscompilations in Smart-Contract Compilers with
  Assertion-Driven Equivalence Oracles}

\author{Alexander Stromberger}
\orcid{0009-0001-6285-0997}
\affiliation{%
  \institution{TU Wien}
  \city{Vienna}
  \country{Austria}}
\email{alexander.stromberger@tuwien.ac.at}

\author{Christoph Hochrainer}
\orcid{0009-0009-4253-4821}
\affiliation{%
  \institution{TU Wien}
  \city{Vienna}
  \country{Austria}}
\email{christoph.hochrainer@tuwien.ac.at}

\author{Valentin W\"ustholz}
\orcid{0000-0003-1496-1104}
\affiliation{%
  \institution{Consensys Diligence}
  \city{Vienna}
  \country{Austria}}
\email{valentin.wustholz@diligence.security}

\author{Maria Christakis}
\orcid{0000-0002-2649-1958}
\affiliation{%
  \institution{TU Wien}
  \city{Vienna}
  \country{Austria}}
\email{maria.christakis@tuwien.ac.at}

%% \renewcommand{\shortauthors}{Trovato et al.}

\begin{abstract}
Smart contracts compiled to the Ethereum Virtual Machine (EVM) can
manage high-value assets and are difficult to patch after deployment.
Compiler miscompilations are therefore especially dangerous:
compilation succeeds, but the generated bytecode does not preserve the
semantics of the source program. Existing smart-contract compiler
fuzzers either focus on robustness failures or rely on differential
oracles that compare compiler configurations.

We present \tool, an approach for finding miscompilations in EVM
smart-contract compilers using assertion-driven equivalence oracles.
\tool uses product programs to encode semantic equivalence as
executable assertions. The key shift is to compile these self-checking
programs and let off-the-shelf fuzzers and symbolic execution engines
explore the resulting bytecode for assertion violations, without
requiring compiler-specific support from the analyzer.

\tool instantiates this idea with two complementary families of
metamorphic transformations. The first, which we call smart-contract
transformations, starts from equivalent expressions and lifts them
into smart-contract code that deliberately exercises EVM-specific
features such as storage, data locations, control flow, contract
interactions, and inline assembly. The second generalizes Equivalence
Modulo Inputs (EMI) from fixed concrete inputs to symbolic input
spaces. Our symbolic EMI records coverage in the original computation
and uses it to guard mutations in the transformed
computation. Together, these transformations produce product programs
that assert that paired computations yield the same outputs.

We evaluate \tool on three EVM smart-contract compilers---\solc,
\solx, and \vyper---using several analysis tools. Overall, \tool
exposes 29 bugs. Of these, 24 are compiler bugs:
6 miscompilations, 17 internal compiler errors, and
1 hang. The remaining 5 are bugs in the analysis
tools. In time-to-bug experiments over 20 known and newly
discovered miscompilations, \tool exposes 9 within 72
hours. Ablations show complementary effects: disabling symbolic EMI
leaves 5 bugs exposed, disabling smart-contract transformations
leaves 4, and disabling both leaves 3.
\end{abstract}

%% \begin{CCSXML}
%%   <ccs2012>
%%    <concept>
%%        <concept_id>10011007.10011074.10011099.10011102.10011103</concept_id>
%%        <concept_desc>Software and its engineering~Software testing and debugging</concept_desc>
%%        <concept_significance>500</concept_significance>
%%        </concept>
%%   </ccs2012>
%% \end{CCSXML}

%% \ccsdesc[500]{Software and its engineering~Software testing and debugging}

%% \keywords{static taint analysis, metamorphic testing, equivalence modulo taint, program analysis testing, information-flow analysis, soundness and precision}

%% \received{20 February 2007}
%% \received[revised]{12 March 2009}
%% \received[accepted]{5 June 2009}

\maketitle

\input{intro}
\input{background}
\input{overview}
\input{contracts}
\input{emi}
\input{eval}
\input{related}
\input{conclusion}

\begin{acks}
We thank the developers of the compilers and analyzers for their
responses to our bug reports. This work has been funded by the Vienna
Science and Technology Fund (WWTF) and the City of Vienna [Grant ID:
  10.47379/ICT22007].
\end{acks}

%\newpage

\bibliographystyle{ACM-Reference-Format}
\bibliography{bibliography}

% \appendix

\end{document}

%% file: intro.tex
%!TEX root = main.tex

% -----------------------------------------------------------
\section{Introduction}
\label{sec:intro}
% -----------------------------------------------------------

Smart contracts targeting Ethereum are compiled to bytecode that runs
on the Ethereum Virtual Machine (EVM). They routinely implement
financial protocols, token systems, bridges, auctions, and governance
mechanisms. Once deployed, a contract is difficult to patch without an
explicit upgrade mechanism, and even upgradeable systems may preserve
buggy state or expose users before a fix can be applied. Compiler
correctness is therefore a security property of the smart-contract
ecosystem: source-level reasoning is only meaningful if the compiler
preserves the semantics of the source program.

\emph{Miscompilations} are particularly dangerous because compilation
succeeds without an obvious failure, yet the generated bytecode does
not preserve the source semantics. A contract that appears correct in
source form may therefore behave differently after compilation.

\textbf{The oracle problem.}
Existing fuzzers and generators for Solidity and related
smart-contract compilers have substantially improved compiler
robustness. \fuzzol~\cite{MitropoulosSotiropoulos2023} uses
syntax-aware AST mutation to exercise new \solc paths;
\erwin~\cite{MaDonaldson2026} uses bounded exhaustive random
generation to target bug-prone Solidity feature combinations; and
\acf~\cite{GroceVanTonder2022} augments AFL with compiler-oriented
mutation and splicing operators. These tools are effective at exposing
robustness failures such as crashes, hangs, and internal compiler
errors.

Semantic testing requires an oracle for determining whether
successfully compiled bytecode preserves the intended behavior. Two
independently developed Solidity compiler-testing tools share the same
name, differing only in capitalization: Li et al.'s
\solsmith~\cite{LiLiu2025} and Shastry's
\efsolsmith~\cite{Shastry2026}. Both use differential
testing~\cite{McKeeman1998} across compiler configurations.
\solsmith~\cite{LiLiu2025} emphasizes validity and language-feature
diversity, whereas \efsolsmith~\cite{Shastry2026} compares legacy with
intermediate-representation-based code generation and unoptimized with
optimized compilation. Because both rely on cross-configuration
agreement, they can miss bugs that affect both sides identically. This
leaves a complementary challenge: detecting semantic errors without
another compiler pipeline or configuration as a reference.

This challenge reflects the oracle problem in compiler testing:
testing tools need a way to distinguish correct generated code from
miscompiled code~\cite{ChenPatra2020}. Notable examples include
\csmith, which generates random C programs and uses differential
testing across compilers~\cite{YangChen2011}, and metamorphic
testing~\cite{ChenCheung1998}, which constructs related programs or
inputs whose executions should satisfy known relations. Equivalence
Modulo Inputs (EMI) is one such technique: it mutates code that is
irrelevant for a given input~\cite{LeAfshari2014}. These techniques
provide powerful semantic oracles. Our goal is to generate equivalence
oracles tailored to EVM-specific language and compiler features and
encode them as assertions that off-the-shelf analyzers can attempt to
violate.

\textbf{Our approach.}
We present \tool, which realizes this idea for EVM smart-contract
compilers. \tool uses product programs to encode semantic equivalence
as executable assertions: it combines computations that should be
equivalent in a single program, executes them on the same inputs, and
compares their results. After compilation, any assertion violation is
a candidate miscompilation.

Embedding the oracle in the program separates oracle construction from
input exploration. A fuzzer or symbolic execution engine need not
understand the compiler, the transformation that produced the
equivalent computations, or why they should agree. It only needs to
explore the compiled program and report assertion violations. Unlike a
differential oracle, each generated product program carries its own
semantic check and does not require another compiler pipeline or
configuration as a reference.

\textbf{Oracle generators.}
\tool constructs assertion-driven equivalence oracles using two
complementary families of metamorphic transformations. The first lifts
equivalent expressions into smart-contract code that exercises
EVM-specific compiler features. These smart-contract transformations
are not generic syntactic perturbations: they deliberately stress
features such as storage, data locations, control flow, contract
interactions, and inline assembly. This allows the generated programs
to reach compiler components that simple expression tests do not
exercise.

The second family introduces symbolic EMI, which generalizes
Equivalence Modulo Inputs from fixed concrete inputs to symbolic input
spaces. Classical EMI executes a program on a concrete input and
mutates code that is irrelevant for that execution. \tool instead
embeds an original and a transformed computation in a single product
program: the first records coverage, and the second uses that coverage
to select between original and mutated code. The product program then
asserts that both computations produce the same outputs.

Together, the two transformation families generate programs that
exercise smart-contract compiler features and check semantic
equivalence. Existing fuzzers and symbolic execution engines can then
search these programs for inputs that expose miscompilations as
assertion violations.

\textbf{Results.}
We evaluate \tool on \solc, \solx, and \vyper using several EVM
analysis tools. \tool finds 24 previously unknown compiler
bugs: 6 miscompilations, 17 internal compiler errors,
and 1 hang. During analyzer-driven exploration, it also exposes
5 bugs in the analysis tools, for 29 bugs overall. The
miscompilations affect multiple compiler components and include bugs
that manifest only after successful compilation. Time-to-bug
experiments and ablations show that the two transformation families
are complementary: using both exposes substantially more bugs than
either family alone.

\textbf{Contributions.}
This paper makes the following contributions:

\begin{itemize}
  \item We introduce assertion-driven equivalence oracles for finding
    miscompilations in EVM smart-contract compilers. \tool encodes
    these oracles in product programs, allowing existing fuzzers and
    symbolic execution engines to search compiled bytecode for inputs
    that expose semantic compiler bugs as assertion violations.

  \item We present smart-contract transformations that lift equivalent
    expressions into code exercising storage, data locations, control
    flow, contract interactions, and inline assembly.

  \item We introduce symbolic EMI, a coverage-guided metamorphic
    formulation of Equivalence Modulo Inputs that makes coverage part
    of the generated program, guards mutations with that coverage, and
    lets existing analyzers search for inputs that violate output
    equivalence.

  \item We implement these ideas in \tool and evaluate them on \solc,
    \solx, and \vyper. \tool finds 24 compiler bugs, including
    6 miscompilations, and exposes 5 additional bugs in
    the analysis tools. Our ablations show that the two transformation
    families are complementary for bug discovery.
\end{itemize}

%%% Local Variables:
%%% mode: latex
%%% TeX-master: "main"
%%% End:

%% file: background.tex
%!TEX root = main.tex

% -----------------------------------------------------------
\section{Background}
\label{sec:background}
% -----------------------------------------------------------

% -----------------------------------------------------------
\subsection{EVM Smart-Contract Compilation}
% -----------------------------------------------------------

Compilers for EVM smart contracts translate high-level contract code
into EVM bytecode. The EVM execution model is stack-based and provides
contract-specific features such as persistent storage, transient
storage, memory, calldata, logs, calls, and exceptional
termination. Correct compilation must therefore preserve not only
arithmetic and control-flow behavior, but also the observable effects
of these EVM-specific features.

The reference Solidity compiler, \solc, exposes multiple compilation
pipelines and optimization settings. In its non-legacy pipeline, \solc
translates Solidity to Yul, a structured intermediate language,
applies most optimization passes at the Yul level, and then lowers the
optimized Yul to EVM bytecode. Yul is also the language used in
Solidity's inline assembly. \solx is an LLVM-based Solidity compiler:
it reuses the \solc frontend but uses LLVM for middle-end and backend
compilation before targeting the EVM. We also consider the \vyper
compiler, which implements a separate smart-contract language and
compiler pipeline targeting the EVM.

We use \emph{miscompilation} for a semantic-preservation failure in
which compilation succeeds but the emitted bytecode behaves
inconsistently with the source-level or intermediate-language
semantics. We distinguish miscompilations from robustness failures
such as crashes, hangs, and internal compiler errors.

% -----------------------------------------------------------
\subsection{Semantic Oracles for Compiler Testing}
% -----------------------------------------------------------

Testing for miscompilations requires a semantic oracle for determining
whether generated code preserves the input program's behavior. Because
the exact expected behavior is often unknown, compiler testing can
instead rely on relations between executions. Differential
testing~\cite{McKeeman1998} compares code produced by different
compilers, versions, pipelines, or configurations, whereas metamorphic
testing generates related inputs whose executions should satisfy a
known relation~\cite{ChenCheung1998}.

In this paper, the central relation is equivalence. For example,
\code{2 * x} and \code{x + x} should evaluate to the same value on the
same input, subject to the arithmetic semantics of the language. A
compiler-testing tool can use such equivalences as oracles: if two
semantically equivalent computations produce different results after
compilation, this indicates a semantic discrepancy.

EMI provides a complementary way to obtain equivalent
programs~\cite{LeAfshari2014}. Given a program and input, statements
not executed on that input can be mutated without changing its
behavior. Classical EMI identifies such input-irrelevant code through
concrete executions and checks that the original and mutated programs
agree on the same input.

%%% Local Variables:
%%% mode: latex
%%% TeX-master: "main"
%%% End:

%% file: overview.tex
%!TEX root = main.tex

% -----------------------------------------------------------
\section{Overview}
\label{sec:overview}
% -----------------------------------------------------------

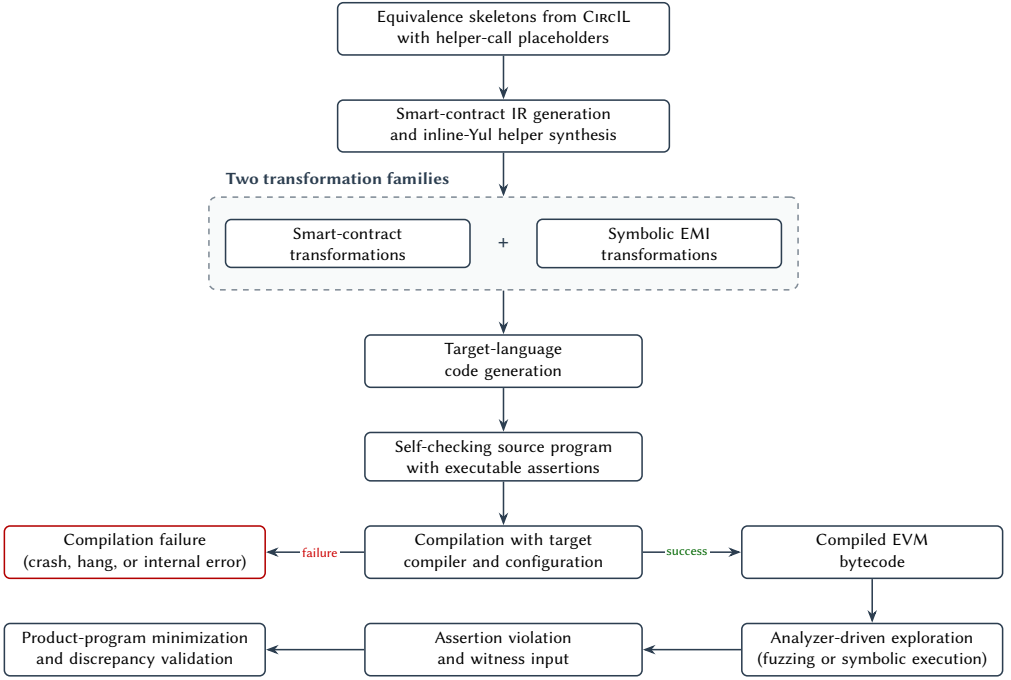
\begin{figure}[t]
  \centering
  \resizebox{0.95\linewidth}{!}{%
  \begin{tikzpicture}[
    >=Latex,
    node distance=8mm and 8mm,
    basebox/.style={
      draw=slateblue,
      thick,
      rounded corners=3pt,
      align=center,
      font=\small\sffamily,
      inner sep=4pt,
      fill=white,
      text width=3.2cm,
      minimum height=8mm
    },
    line/.style={
      ->,
      >=Stealth,
      thick,
      draw=slateblue
    },
    labelstyle/.style={
      font=\scriptsize\sffamily,
      fill=white,
      inner sep=1pt
    },
    group/.style={
      draw=slateblue!60,
      dashed,
      thick,
      rounded corners=4pt,
      fill=lightslate!50
    },
    plus/.style={
      font=\Large\sffamily\bfseries,
      text=slateblue
    }
  ]

  % --- CENTRAL AXIS ---
  \coordinate (axis) at (0,0);

  % --- EQUIVALENCE AND PROGRAM CONSTRUCTION ---
  \node[
    basebox,
    text width=5.8cm,
    anchor=north
  ] (circil) at (axis)
    {Equivalence skeletons from \circil\\
     with helper-call placeholders};

  \node[
    basebox,
    text width=5.8cm,
    below=8mm of circil
  ] (construction)
    {Smart-contract IR generation\\
     and inline-Yul helper synthesis};

  % --- COMPOSED TRANSFORMATION FAMILIES ---
  \node[
    group,
    minimum width=10.8cm,
    minimum height=17mm,
    below=8mm of construction
  ] (transformations) {};

  \node[
    basebox,
    text width=4.2cm,
    anchor=east,
    xshift=-6mm
  ] (sct) at (transformations.center)
    {Smart-contract\\transformations};

  \node[
    plus
  ] (plus) at (transformations.center)
    {+};

  \node[
    basebox,
    text width=4.2cm,
    anchor=west,
    xshift=6mm
  ] (semi) at (transformations.center)
    {Symbolic EMI\\transformations};

  \node[
    font=\small\sffamily\bfseries,
    text=slateblue,
    anchor=south west,
    xshift=2mm,
    yshift=1mm
  ] at (transformations.north west)
    {Two transformation families};

  % --- SOURCE-PROGRAM GENERATION ---
  \node[
    basebox,
    text width=4.8cm,
    below=8mm of transformations
  ] (codegen)
    {Target-language\\code generation};

  \node[
    basebox,
    text width=4.8cm,
    below=8mm of codegen
  ] (product)
    {Self-checking source program\\
     with executable assertions};

  \node[
    basebox,
    text width=4.8cm,
    below=8mm of product
  ] (compile)
    {Compilation with target\\
     compiler and configuration};

  % --- COMPILATION OUTCOMES ---
  \node[
    basebox,
    left=18mm of compile,
    text width=4.5cm,
    draw=red!70!black
  ] (robust)
    {Compilation failure\\
     (crash, hang, or internal error)};

  \node[
    basebox,
    right=18mm of compile,
    text width=4.5cm
  ] (bytecode)
    {Compiled EVM\\bytecode};

  % --- ANALYSIS PIPELINE ---
  \node[
    basebox,
    text width=4.5cm,
    below=8mm of bytecode
  ] (analyzer)
    {Analyzer-driven exploration\\
     (fuzzing or symbolic execution)};

  \node[
    basebox,
    text width=4.8cm,
    at={(compile.south |- analyzer.north)},
    anchor=north
  ] (violation)
    {Assertion violation\\
     and witness input};

  \node[
    basebox,
    text width=4.5cm,
    at={(robust.south |- analyzer.north)},
    anchor=north
  ] (validate)
    {Product-program minimization\\
     and discrepancy validation};

  % --- ROUTING ---
  \draw[line] (circil) -- (construction);
  \draw[line] (construction) -- (transformations);
  \draw[line] (transformations) -- (codegen);
  \draw[line] (codegen) -- (product);
  \draw[line] (product) -- (compile);

  \draw[line]
    (compile)
    --
    node[
      labelstyle,
      pos=0.45,
      text=red!80!black
    ] {failure}
    (robust);

  \draw[line]
    (compile)
    --
    node[
      labelstyle,
      pos=0.45,
      text=green!40!black
    ] {success}
    (bytecode);

  \draw[line] (bytecode) -- (analyzer);
  \draw[line] (analyzer) -- (violation);
  \draw[line] (violation) -- (validate);

  \end{tikzpicture}%
  }
\caption{Overview of the \tool workflow. \circil supplies equivalence
  skeletons, which \tool translates into a smart-contract IR. \tool
  may apply smart-contract transformations and, where applicable,
  symbolic EMI before emitting self-checking source programs and
  compiling them for analyzer-driven exploration.}
  \label{fig:overview-workflow}
\end{figure}

\figref{fig:overview-workflow} summarizes the end-to-end \tool
workflow. Reading the figure from top to bottom, \tool generates a
self-checking source program from equivalent expressions, compiles it,
and passes the resulting bytecode to an off-the-shelf analyzer. This
separates generation of the semantic checks from exploration of the
program's input space.
The figure shows the compiler-testing path. Analyzer bugs encountered
during exploration are recorded separately and discussed in
\secref{sec:evaluation}.

\textbf{Equivalence skeletons.}
\tool begins with equivalent expressions generated using \circil, an
expression-level intermediate representation introduced in prior
work~\cite{HochrainerIsychev2025}. \circil represents
language-independent computations and generates expression
equivalences that hold by construction. A generated equivalence may
contain only ordinary expressions, or it may include placeholders for
helper computations. When such a placeholder is present, \tool
realizes it as a pure Solidity helper function whose side-effect-free
body is generated directly in Yul, Solidity's structured low-level
language:

\begin{center}
\code{yul\_func(x) + yul\_func(x) == 2 * yul\_func(x)}.
\end{center}

\noindent
Here, \code{yul\_func(x)} denotes the same helper computation at each
occurrence. \circil generates the surrounding algebraic equivalence
but not the helper body, which \tool generates later. Equivalences
without helper-call placeholders proceed through the pipeline without
a generated Yul helper.

\begin{figure}[t]
\begin{lstlisting}[language=Solidity]
contract C {
  function yul_func(uint256 x) internal pure returns (uint256 r) {
    assembly {
      r := add(x, 1)
      if lt(x, 10) {
        r := add(r, 2)
      }
    }
  }

  function check(uint256 x) public pure {
    unchecked {
      assert(yul_func(x) + yul_func(x) == 2 * yul_func(x));
    }
  }
}
\end{lstlisting}
\vspace{-1em}
\caption{Schematic smart-contract IR generation for an equivalence
  containing a helper-call placeholder. \tool places the generated
  Yul helper and surrounding equivalence check in an enclosing
  Solidity contract.}
\vspace{-1em}
\label{fig:overview-ir-example}
\end{figure}

\textbf{Smart-contract IR generation.}
\tool translates the expression skeleton into an internal
representation (IR) for smart contracts. This IR adds the enclosing
contract structure and represents the functions, statements, types,
and equivalence checks that make up the generated program.
When an equivalence contains a helper-call placeholder, \tool
synthesizes a corresponding function whose body is generated directly
in Yul. When no placeholder is present, this synthesis step is
skipped. \figref{fig:overview-ir-example} shows the schematic Solidity
form of the running example.

In this example, the generated Yul helper first computes \code{x +
  1}. When \code{x < 10}, it additionally adds \code{2}.  Each call
evaluates the same side-effect-free computation on the same argument,
so all three calls to \code{yul\_func(x)} return the same value. The
assertion therefore instantiates the identity \code{v + v == 2 * v},
where \code{v} is the value returned by the helper. The
\code{unchecked} block specifies the arithmetic semantics of the
surrounding Solidity expression. Solidity normally reverts when
\code{uint256} addition or multiplication overflows, whereas Yul and
the EVM use arithmetic modulo \(2^{256}\). Placing the assertion in an
\code{unchecked} block makes the addition and multiplication in the
assertion use the same modular semantics. The equivalence therefore
remains valid for every \code{uint256} input.

\textbf{Transformation families.}
\tool supports two complementary families of metamorphic
transformations. In its main configuration, \tool applies
smart-contract transformations and, when a suitable Yul helper is
present, also applies symbolic EMI. Smart-contract transformations
operate on the enclosing smart-contract program, whereas symbolic EMI
requires a side-effect-free computation that can be duplicated and
mutated. Since generated Yul helpers satisfy this requirement,
symbolic EMI is applied to them; when no suitable helper is present,
the symbolic-EMI stage is skipped. The running example illustrates a
program to which both transformation families are applied.

\begin{figure}[t]
\begin{lstlisting}[language=Solidity]
function check(uint256 x) public pure {
  uint256[1] memory factor;
  factor[0] = 2;

  unchecked {
    assert(yul_func(x) + yul_func(x) == factor[0] * yul_func(x));
  }
}
\end{lstlisting}
\vspace{-1em}
\caption{Schematic smart-contract transformation. The
  constant-to-local rewrite replaces the literal \code{2} with an
  equivalent read from a memory array while leaving the generated Yul
  helper unchanged.}
\vspace{-1em}
\label{fig:overview-smart-contract-transformation}
\end{figure}

\textbf{Smart-contract transformations.}
Smart-contract transformations lift expression-level equivalences into
enclosing code that exercises smart-contract-specific compiler
features. They modify the surrounding contract while leaving generated
Yul helper bodies unchanged. Symbolic EMI instead transforms these
generated helpers directly.

\figref{fig:overview-smart-contract-transformation} illustrates a
constant-to-local rewrite applied to the running example. The
transformation introduces a one-element memory array, stores \code{2}
in it, and replaces the literal \code{2} in the assertion with the
read \code{factor[0]}. Because this read yields the same value, the
asserted equivalence is preserved. The program nevertheless exercises
memory allocation, array indexing, and data-location handling. Other
smart-contract transformations introduce storage, control flow,
function boundaries, contract interactions, and related language
features. They thereby increase compiler-feature coverage without
changing the equivalence checked by the generated program.

\textbf{Symbolic EMI transformations.}
When symbolic EMI is applied to a generated Yul helper, \tool selects
one or more assignments and derives mutated alternatives for their
right-hand-side expressions. In the running example, \tool selects the
conditional assignment \code{r := add(r, 2)} and derives the mutant
\code{add(add(r, 2), 1)} from its right-hand side. The mutated
expression is deliberately different from the original expression.
The transformed computation uses the mutation only when coverage
indicates that the original computation did not reach the selected
assignment.

\tool rewrites the helper to contain an original and a transformed
computation. The original computation records whether it reaches the
selected assignment. At the corresponding point, the transformed
computation evaluates the original expression when the assignment was
covered and the mutated expression otherwise. Because the generated
Yul helper is side-effect free, \tool can evaluate both computations
during one invocation and compare their results without duplicating
observable effects. \figref{fig:overview-symbolic-emi} illustrates
this construction.

For an input satisfying \code{x < 10}, the original computation
executes the selected assignment, produces \code{x + 3}, and then
sets \code{covered} to one. The transformed computation enters the same
branch and, because \code{covered} is one, evaluates the original
right-hand side and produces the same result. Otherwise, the
original computation does not reach the selected assignment and
\code{covered} remains zero. The transformed computation also skips
the branch, so the mutated expression is not evaluated and cannot
affect the result. This is the EMI relation: the mutation is
irrelevant for that input, but its presence can still expose incorrect
optimization or code-generation behavior.

The helper finally compares \code{original} and \code{transformed} and
reverts if they differ. Under the intended semantics, the comparison
holds for every value of \code{x}. Because coverage is computed within
the generated program, different inputs explored by a fuzzer or
symbolic execution engine induce their own coverage information
without requiring the oracle to be regenerated. Symbolic EMI thereby
generalizes classical EMI from a fixed concrete input to a symbolic
input space.

Symbolic EMI could in principle be applied directly to Solidity.
\tool instead applies it to generated Yul helpers because Yul is
compact and structured, allowing coverage instrumentation and
expression mutation to be defined uniformly without handling
Solidity's broader high-level semantics. This choice also targets a
central optimization layer in \solc: in the non-legacy pipeline,
Solidity is translated to Yul, most optimization passes operate on Yul,
and the optimized Yul is then lowered to EVM bytecode. Symbolic EMI
therefore exercises both Yul optimization and subsequent Yul-to-EVM
code generation.

\begin{figure}[t]
\begin{lstlisting}[language=Solidity]
function yul_func(uint256 x) internal pure returns (uint256 r) {
  assembly {
    let covered := 0

    // Original computation
    let original := add(x, 1)
    if lt(x, 10) {
      original := add(original, 2)
      covered := 1      
    }

    // Transformed computation
    let transformed := add(x, 1)
    if lt(x, 10) {
      if covered {
        transformed := add(transformed, 2)
      }
      if iszero(covered) {
        transformed := add(add(transformed, 2), 1)
      }
    }

    // Equivalence check
    if iszero(eq(original, transformed)) {
      revert(0, 0) // represents assert(false)
    }
    r := original
  }
}
\end{lstlisting}
\vspace{-1em}
\caption{Schematic symbolic-EMI transformation of the generated Yul
  helper \code{yul\_func}. The first computation records whether it
  reaches the selected assignment. At the corresponding point, the
  second uses the original expression when the assignment was covered
  and a mutated expression otherwise.}
\vspace{-1em}
\label{fig:overview-symbolic-emi}
\end{figure}

\textbf{Self-checking source generation.}
After applying the selected transformations, \tool emits a
self-checking source program whose executable checks encode semantic
equivalences. The \circil-derived check compares computations that are
equivalent by construction. When symbolic EMI is applied, an additional
check compares the original and transformed Yul computations. Both are
instances of a \emph{product-program} construction: related
computations are combined in one program, and their results are
compared within that program. Product constructions reduce relational
reasoning about multiple computations to an ordinary safety
property~\cite{ClarksonSchneider2008,BartheDArgenio2004,TerauchiAiken2005,BartheCrespo2011}.

Product programs are an established technique. \tool uses them to
expose semantic equivalence as executable checks that off-the-shelf
fuzzers and symbolic execution engines can explore systematically.
Under the intended semantics, these checks should hold for every input
on which the related computations are defined.

\textbf{Compilation and exploration.}
\tool compiles each self-checking source program with the selected
target compiler and configuration. A compiler crash, hang, or internal
compiler error is recorded as a robustness failure. When compilation
succeeds, the resulting EVM bytecode is passed to a fuzzer or symbolic
execution engine. The analyzer need not understand \circil, the
smart-contract IR, the applied transformations, or the target compiler.
It only needs to execute the bytecode and search for an input that
causes one of the encoded equivalence checks to fail.

Analyzer crashes are recorded separately as analyzer bugs. If an
analyzer reports an assertion violation that does not reproduce under
concrete execution, we classify it as an analyzer-imprecision bug.
Otherwise, \tool minimizes the self-checking program and validates the
discrepancy against the intended source- or intermediate-level
semantics. The absence of a reported violation does not establish
compiler correctness; it means only that the analyzer found no
counterexample within the explored input space.

%%% Local Variables:
%%% mode: latex
%%% TeX-master: "main"
%%% End:

%% file: contracts.tex
%!TEX root = main.tex

\begin{table*}[t]
\caption{Smart-contract transformations implemented by \tool.}
\vspace{-1em}  
\label{tab:smart-contract-transformations}
\centering
\small
\begin{tabular}{p{0.23\textwidth}p{0.72\textwidth}}
\toprule
\textbf{Transformation} & \textbf{Rewrite}\\
\midrule
Constant-to-state rewrite
  & Replace a constant with an equivalent access through contract
    state.\\
Constant-to-local rewrite
  & Replace a constant with an equivalent access through local
    memory.\\
Local-to-state rewrite
  & Replace a local variable with an equivalent state variable.\\
Struct wrapping
  & Represent a value through an equivalent struct-field access.\\
Enum wrapping
  & Encode an unsigned integer as an enum and recover its value.\\
Storage-layout variation
  & Change the contract's base storage slot.\\
State-variable insertion
  & Insert unused state variables to perturb the storage layout.\\
\midrule
Ternary lowering
  & Replace a ternary expression with equivalent explicit control
    flow.\\
Expression extraction
  & Break nested subexpressions into temporary assignments.\\
Local-variable reuse
  & Reuse a same-typed local variable after its live range ends.\\
Snippet injection
  & Insert inline Yul without changing the surrounding computation.\\
Event emission
  & Emit events using constants and local values.\\
\midrule
Function extraction
  & Extract a subexpression into a function and replace it with a
    call.\\
Contract extraction
  & Move an eligible function and its transitive dependencies to a
    new contract or library and replace its calls with external
    calls.\\
Try/catch wrapping
  & Route an external contract call through an equivalent try/catch
    wrapper.\\
\bottomrule
\end{tabular}
\vspace{-1em}
\end{table*}

% -----------------------------------------------------------
\section{Smart-Contract Transformations}
\label{sec:contracts}
% -----------------------------------------------------------

This section presents \tool's smart-contract transformations. We first
describe their design and the implemented transformations, then discuss
equivalence preservation and composition, and finally illustrate their
interaction with a representative \solc miscompilation.

% -----------------------------------------------------------
\subsection{Transformation Design}
\label{sec:smart-contract-lifting}
% -----------------------------------------------------------

Once \tool has translated an expression-level equivalence into its
smart-contract IR, it applies transformations to the enclosing
program. Each transformation changes how values are represented, how
computations are structured, or the context in which they execute,
while preserving the values compared by the generated assertion.
Generated Yul helper bodies remain unchanged.

\tabref{tab:smart-contract-transformations} summarizes the
transformations.
The transformations
fall into three categories, separated by horizontal rules. The first
group changes the representation or location of values. For example,
constant-to-state and constant-to-local rewrites replace literals with
equivalent accesses through state or local memory, while enum and
struct wrapping represent values through additional Solidity types.
This group also includes local-to-state rewriting, storage-layout
variation, and insertion of unused state variables to perturb the
contract storage layout.

The second group changes the local structure of computations or
introduces auxiliary statements within functions. Ternary lowering
replaces a ternary expression with equivalent explicit control flow,
while expression extraction breaks nested subexpressions into
temporary assignments. Local-variable reuse changes how intermediate
values are stored, snippet injection introduces additional inline Yul,
and event emission inserts additional statements without changing the
values checked by the assertion.

The final group changes function and contract structure. Function
extraction replaces a selected expression \(e\) with a call to a fresh
function parameterized by the variables referenced by \(e\). Contract
extraction moves an eligible function, together with its transitive
dependencies, into a new contract or library and replaces its calls
with external calls. Try/catch wrapping introduces an additional
boundary around external contract calls.

Function and contract extraction are applied early in the
transformation pipeline, before later transformations introduce
constructs whose behavior may depend on the enclosing contract. Later
transformations may then operate on structures introduced by these
earlier stages. This composition allows a simple expression-level
equivalence to exercise several interacting smart-contract features
in the same generated program.

% -----------------------------------------------------------
\subsection{Equivalence Preservation and Composition}
\label{sec:smart-contract-composition}
% -----------------------------------------------------------

Each smart-contract transformation preserves the values compared by
the generated assertion under the intended semantics. A value may, for
example, be moved from a local variable to state, a computation may be
routed through additional control flow or function calls, and
auxiliary constructs may introduce effects unrelated to the checked
values. \tool applies a transformation only when its preconditions
guarantee that these changes cannot affect the equivalence check.

This preservation condition makes the transformations compositional.
Suppose the assertion is valid before applying a transformation. Since
the transformation preserves the compared values, the assertion
remains valid afterward. Repeating this argument for the sequence of
transformations applied by \tool establishes that their composition
preserves the original equivalence. Later transformations may
therefore operate on structures introduced by earlier ones without
requiring a new oracle.

% -----------------------------------------------------------
\subsection{Representative Miscompilation}
\label{sec:smart-contract-example}
% -----------------------------------------------------------

\figref{fig:smart-contract-miscompilation} shows how two transformations
from \tabref{tab:smart-contract-transformations}, contract extraction
and a constant-to-state rewrite, reproduce \code{SignedImmutables}, a
previously known \solc miscompilation documented in Solidity's
official list of known bugs~\cite{SolidityKnownBugs}. Before applying
smart-contract transformations, \tool generates an assertion in
\code{C.f} that compares two calls to the same \code{yul\_fragment}
helper. Contract extraction moves the second helper into contract
\code{D}, so \code{C.f} compares \code{C.yul\_fragment()} with
\code{D.yul\_fragment()}. The constant-to-state rewrite then replaces
the local constant in \code{D.yul\_fragment} with the immutable state
variable \code{D.i}, initialized to the same value.

The final program therefore computes the same Yul expression in two
different smart-contract contexts. In \code{C.yul\_fragment},
\code{l} is initialized directly to the \code{int16} value
\code{-8283}. In \code{D.yul\_fragment}, \code{l} receives the same
value through \code{D.i}. Both functions then evaluate the identical
Yul comparison \code{lt(l, 0xe3e3)}, so they should return the same
result and the assertion in \code{C.f} should hold.

The affected compiler, however, did not properly sanitize signed
immutable values shorter than 256 bits before they were accessed from
inline assembly. Consequently, the value observed by
\code{D.yul\_fragment} had different higher-order bits from the value
observed by \code{C.yul\_fragment}. The comparison in
\code{C.yul\_fragment} evaluated to false, whereas the same comparison
in \code{D.yul\_fragment} evaluated to true, causing the assertion in
\code{C.f} to fail.

This example illustrates why the transformations are designed to
compose. Neither contract extraction nor the constant-to-state rewrite
changes the intended result. Together, however, they place equivalent
computations on different compilation paths involving contract
creation, immutable state, and inline assembly, exposing a semantic
discrepancy that is absent from the original expression-level
equivalence.

\begin{figure}[t]
\begin{lstlisting}[language=Solidity]
contract D {
  int16 private immutable i = int16(-8283);

  function yul_fragment() external view returns (bool r) {
    int16 l = i;
    assembly {
      r := lt(l, 0xe3e3)
    }
  }
}

contract C {
  function yul_fragment() internal view returns (bool r) {
    int16 l = -8283;
    assembly {
      r := lt(l, 0xe3e3)
    }
  }

  function f() public {
    assert(yul_fragment() == (new D()).yul_fragment());
  }
}
\end{lstlisting}
\vspace{-1em}
\caption{Program reproducing the known \code{SignedImmutables}
  miscompilation in \solc. In \code{C.yul\_fragment}, the value
  \code{-8283} is stored directly in a local variable; in
  \code{D.yul\_fragment}, the same value is obtained from the
  immutable state variable \code{D.i}.}
\vspace{-1em}
\label{fig:smart-contract-miscompilation}
\end{figure}

%%% Local Variables:
%%% mode: latex
%%% TeX-master: "main"
%%% End:

%% file: emi.tex
%!TEX root = main.tex

% -----------------------------------------------------------
\section{Symbolic EMI}
\label{sec:emi}
% -----------------------------------------------------------

This section presents symbolic EMI, which extends Equivalence Modulo
Inputs from a fixed concrete input to an input space explored by a
fuzzer or symbolic execution engine. We first explain this
generalization, then describe the coverage-guided transformation and
its correctness argument, and finally show how symbolic EMI exposes a
representative miscompilation.

% -----------------------------------------------------------
\subsection{From EMI to Symbolic EMI}
\label{sec:symbolic-emi-motivation}
% -----------------------------------------------------------

Classical EMI ties each transformed program to a concrete input and
the coverage observed on that input~\cite{LeAfshari2014}. Symbolic EMI
instead embeds this coverage relation into the generated program, so a
single program realizes the corresponding EMI relation for every input
explored by the analyzer. Here, \emph{symbolic} refers to this
generalization from a fixed concrete input to an input space; it does
not require symbolic execution.

\tool applies this construction to side-effect-free Yul computations.
We next define the coverage-guided transformation independently of the
running example from \secref{sec:overview}.

% -----------------------------------------------------------
\subsection{Coverage-Guided Transformation}
\label{sec:symbolic-emi-transformation}
% -----------------------------------------------------------

\begin{algorithm}[t]
\caption{Symbolic-EMI transformation}
\label{alg:symbolic-emi}
\small
\KwIn{Side-effect-free computation \(P\)}
\KwOut{Product computation \(Q\)}

\(P_o \gets\) copy of \(P\)\;
\(P_t \gets\) copy of \(P\)\;
Rename variables declared in \(P_t\) to fresh variables\;
Select at random one to three assignments in \(P\) as \(T\)\;

\For{each target assignment \(x_i := e_i \in T\)}{
  Randomly derive a mutant \(e_i'\) from \(e_i\)\;
  Introduce a coverage flag \(c_i \gets \mathsf{false}\)\;
  Replace \(x_i := e_i\) in \(P_o\) with
  \(x_i := e_i;\ c_i \gets \mathsf{true}\)\;
  Replace the corresponding assignment in \(P_t\) with one using its
  original right-hand side if \(c_i\), and a mutant otherwise\;
}

\(Q \gets P_o;\ P_t\)\;
Append to \(Q\) an assertion that the outputs of \(P_o\) and \(P_t\)
are equal\;
\Return \(Q\)\;
\end{algorithm}

\algoref{alg:symbolic-emi} summarizes the symbolic-EMI transformation.
Given a side-effect-free Yul computation \(P\), \tool constructs two
copies of \(P\), \(P_o\) and \(P_t\), and renames the variables
declared in \(P_t\) so that the copies can coexist in one product
computation. \tool selects at random one to three assignments as
mutation targets \(T\).

For each target \(x_i := e_i\), \tool derives a mutant \(e_i'\) while
retaining \(e_i\) in the original computation. It randomly selects one
of six mutation operators: applying \code{not} or \code{iszero},
adding, XORing, or subtracting a random nonzero constant, or replacing
the expression with a random constant. \tool also introduces a
coverage flag \(c_i\), which \(P_o\) sets after executing the target.
The corresponding assignment in \(P_t\) uses the original expression
when \(c_i\) is set and its mutant otherwise. Finally, \tool combines
\(P_o\) and \(P_t\) and adds an assertion comparing their outputs.
Coverage is tracked independently for each target, allowing a single
product program to contain up to three guarded mutations.

Crucially, the coverage information is produced during execution
rather than for a fixed input in advance. Each input explored by the
analyzer therefore induces its own coverage and, consequently, its own
choice between original and mutated expressions. A single product
program thus realizes the corresponding EMI relation across the input
space explored by the analyzer.

% -----------------------------------------------------------
\subsection{Correctness of the Transformation}
\label{sec:symbolic-emi-correctness}
% -----------------------------------------------------------

Suppose, for contradiction, that the transformed computation evaluates
a mutated expression for some target in $T$, and consider the first
such evaluation across all targets. Until that point, the original and
transformed computations have evaluated corresponding statements using
the same expressions and therefore have identical values for
corresponding computation variables. They consequently follow the same
control flow. If the transformed computation reaches this target
assignment, the original computation must have reached the
corresponding assignment as well. Its coverage flag is therefore set,
causing the transformed computation to evaluate the original rather
than the mutated expression, a contradiction.

Thus, under the intended semantics, no mutated expression is evaluated.
The original and transformed computations consequently produce the
same output for every input on which they terminate normally. A
disagreement after compilation therefore indicates that the compiled
program does not preserve this equivalence, providing the semantic
oracle used by \tool.

% -----------------------------------------------------------
\subsection{Representative Miscompilation}
\label{sec:symbolic-emi-example}
% -----------------------------------------------------------

\begin{figure}[t]
\begin{lstlisting}[language=Solidity]
contract C {
  function f() external {
    assembly {
      // Initialize coverage and memory
      sstore(0xF1a9, 0x00)
      mstore(0x00, 1)
    }

    assembly {
      // Original computation and coverage recording
      let x := mload(0x00)
      sstore(0xF1a9, 0x01)

      // Restore memory before the transformed computation
      mstore(0x00, 1)

      // Transformed computation
      let y
      if sload(0xF1a9) {
        y := mload(0x00)
      }
      if iszero(sload(0xF1a9)) {
        y := add(mload(0x00), 0x42)
      }

      // Equivalence check
      if iszero(eq(x, y)) {
        revert(0, 0) // represents assert(false)
      }
    }
  }
}
\end{lstlisting}
\vspace{-1em}
\caption{Symbolic-EMI program reproducing the known
  \code{InlineAssemblyMemorySideEffects} miscompilation in \solc. The
  storage slot \code{0xF1a9} carries the coverage flag across the two
  inline assembly blocks.}
\vspace{-1em}
\label{fig:symbolic-emi-miscompilation}
\end{figure}

\figref{fig:symbolic-emi-miscompilation} shows how \tool reproduces
\code{InlineAssemblyMemorySideEffects}, a previously known \solc
miscompilation documented in Solidity's official list of known
bugs~\cite{SolidityKnownBugs}. The program contains two inline
assembly blocks. The first initializes the state needed by the
symbolic-EMI construction: it clears the coverage flag stored at
\code{0xF1a9} and writes \code{1} to memory address \code{0x00}.

The second block first executes the original computation. Its selected
target is the assignment \code{x := mload(0x00)}, which reads the
initialized memory value into \code{x}. After executing this
assignment, the program records that the target was reached by setting
the coverage flag to one. A storage slot is used for this flag because
a Yul local variable cannot span the two inline assembly blocks.

Before executing the transformed computation, the program restores
\code{1} at memory address \code{0x00}, so both computations should
observe the same memory state. The transformed computation then
consults the coverage flag. Because the original computation reached
the target, the flag is set and the transformed computation evaluates
the same expression, \code{mload(0x00)}, assigning its result to
\code{y}. Had the target not been reached, it would instead evaluate
the mutant \code{add(mload(0x00), 0x42)}.

Under the intended semantics, both memory loads therefore return
\code{1}, yielding \code{x = y = 1}. The final equivalence check
compares these values and should succeed.

The affected compiler, however, optimized the first inline assembly
block in isolation. Because that block did not itself read the value
written by \code{mstore(0x00, 1)}, the optimizer incorrectly removed
the store without accounting for the read in the following block.
The original computation consequently observes \code{x = 0}. The
later \code{mstore}, which restores memory before the transformed
computation, remains, so the transformed computation observes
\code{y = 1}. The resulting disagreement causes the equivalence check
to fail.

Notably, the mutant is never evaluated in the failing execution.
Instead, the miscompilation changes the memory state observed by two
computations that should agree. This example shows that symbolic EMI
can expose a compiler bug even when the guarded mutant itself is not
executed.

%%% Local Variables:
%%% mode: latex
%%% TeX-master: "main"
%%% End:

%% file: eval.tex
%!TEX root = main.tex

% -----------------------------------------------------------
\section{Experimental Evaluation}
\label{sec:evaluation}
% -----------------------------------------------------------

\begin{table*}[t]
  \centering
  \small
  \caption{Compiler bugs found by \tool.  MC = miscompilation; ICE =
    internal compiler error; H = hang.  Linked bug numbers point to
    public reports, and linked fix letters point to public fix pull
    requests.}
  \vspace{-1em}
  \label{tab:bugs}
  \begin{tabularx}{\textwidth}{clclX}
    \toprule
    \textbf{Bug} & \textbf{System} & \textbf{Type} &
    \textbf{Status} & \textbf{Description} \\
    \midrule

    \bug{1} & \solx & MC & fixed (\fix{A}) &
      EVM stack corruption during branch folding. \\
    \bug{2} & \solx & MC & fixed (\fix{B}) &
      Invalid opcode from deep-stack lowering. \\
    \bug{3} & \solx & MC & fixed (\fix{C}) &
      LLVM \tcode{srem} mismatch for \tcode{INT\_MIN \% -1}. \\
    \bug{4} & \solx & MC & fixed (\fix{D}) &
      Stale EVM stack value after branch elimination. \\
    \bug{5} & \solx & ICE & confirmed &
      Backward-propagation stackification fails. \\
    \bug{6} & \solx & ICE & fixed (\fix{E}) &
      Typed-stack tracking fails across \tcode{SWAPX}/\tcode{DUPX}. \\
    \bug{7} & \solx & H & fixed (\fix{F}) &
      LLVM optimization hangs on irreducible control flow. \\
    \bug{8} & \solx & ICE & fixed (\fix{G}) &
      Invalid \tcode{CALLDATALOAD} rematerialization. \\
    \bug{9} & \solx & ICE & fixed (\fix{H}) &
      Invalid ELF class crashes the linker. \\

    \midrule

    \bug{10} & \solc & MC & confirmed (\fix{I}) &
      Inconsistent dirty-bit handling across compiler pipelines. \\
    \bug{11} & \solc & MC & confirmed &
      Invalid library placeholders in SSA-CFG. \\
    \bug{12} & \solc & ICE & confirmed &
      Oversized EOF bytecode triggers an ICE. \\
    \bug{13} & \solc & ICE & fixed (\fix{J}) &
      Expression simplification produces invalid Yul. \\
    \bug{14} & \solc & ICE & confirmed &
      Stack-layout construction fails to converge. \\
    \bug{15} & \solc & ICE & fixed (\fix{K}) &
      SSA-CFG stack handling violates a \tcode{StackUtils} invariant. \\
    \bug{16} & \solc & ICE & confirmed &
      SSA-CFG layout violates a
      \tcode{StackLayoutGenerator} invariant. \\

    \midrule

    \bug{17} & \vyper & ICE & fixed (\fix{L}) &
      Undefined \tcode{BOT} bound in Venom value-range analysis. \\
    \bug{18} & \vyper & ICE & fixed (\fix{M}) &
      Inconsistent stack model during Venom reordering. \\
    \bug{19} & \vyper & ICE & fixed (\fix{N}) &
      Dead liveness value mishandled during Venom stack cleanup. \\
    \bug{20} & \vyper & ICE & confirmed &
      IR optimization drops a required symbol. \\
    \bug{21} & \vyper & ICE & confirmed &
      Symbol-reference encoding violates an assembler invariant. \\
    \bug{22} & \vyper & ICE & fixed (\fix{O}) &
      Stale liveness after Venom branch optimization. \\
    \bug{23} & \vyper & ICE & confirmed &
      Recursive CFG traversal exceeds the recursion limit. \\
    \bug{24} & \vyper & ICE & confirmed &
      Memory-copy elision assumes every instruction has operands. \\

    \bottomrule
  \end{tabularx}
\end{table*}

We evaluate \tool on \solc, \solx, and \vyper. In our evaluation, we
address the following research questions:

\begin{description}
  \item[RQ1:] How effective is \tool in detecting bugs in EVM
    smart-contract compilers?
  \item[RQ2:] What characterizes the fixed miscompilations found by
    Soliduzz?
  \item[RQ3:] How efficiently does \tool expose known and newly
    discovered miscompilations?
  \item[RQ4:] How do smart-contract transformations and symbolic EMI
  affect bug discovery?
\end{description}

% -----------------------------------------------------------
\subsection{Setup}
\label{sec:setup}
% -----------------------------------------------------------

\textbf{Compiler selection and configuration.}
We evaluate \tool on three EVM compilers: \solc, \solx, and \vyper. We
randomize the target EVM version (Istanbul or later), the optimizer
run count (\code{1}, \code{10}, \code{100}, \code{1000}, \code{10000},
or \code{100000}), and, where supported, whether to use IR, optimize Yul, or optimize
bytecode. We also vary compiler-specific optimization configurations,
including Yul optimizer sequences and optimization pipelines for
\solc, optimizer modes for \solx, and code-generation and optimization
settings for \vyper. Because \vyper does not use Yul, symbolic EMI is
not applied when testing \vyper.

\textbf{Analysis tools.}
For successfully compiled programs, we use the fuzzers
\textsc{Echidna}~\cite{GriecoSong2020} 2.2.7,
\textsc{Forge}~\cite{Foundry} 1.7.1, and
\textsc{Medusa}~\cite{Medusa} 1.5.1; the symbolic execution engine
\textsc{hevm}~\cite{DxoSoos2024} 0.58.0; and \textsc{evmfuzz}, a
black-box fuzzer developed as part of \tool. We use these tools to
explore the resulting EVM bytecode for violations of the embedded
equivalence checks. We do not use \textsc{Medusa} with \solx because
of compatibility issues.

\textbf{Hardware.}
Experiments run on Debian GNU/Linux 13 (Trixie) on a server with two
AMD EPYC 9654 96-core processors and 1.5\,TB of memory.

% -----------------------------------------------------------
\subsection{Results}
\label{sec:results}
% -----------------------------------------------------------

\textbf{RQ1: Bugs.}
We conducted open-ended bug-finding campaigns while developing
\tool. \tabref{tab:bugs} summarizes the resulting compiler bugs,
including their type, status, and a brief description. In total, \tool
found 24 previously unknown compiler bugs across the three
compilers: 6 miscompilations, 17 internal compiler
errors, and 1 hang. All 24 bugs have been confirmed,
14 have already been fixed, and bug~\bug{10} has a proposed fix
that is still pending. For each bug, we followed the developers'
preferred disclosure process. To preserve anonymity during
double-blind review, the public bug reports linked from
\tabref{tab:bugs} were filed from an anonymous account.

\begin{wraptable}{r}{0.45\textwidth}
  \vspace{-1.2em}
  \centering
  \small
  \caption{Bugs in analyzers exposed during the evaluation.  Linked
    bug numbers point to public reports, and linked fix letters point
    to public fixes.}
  \vspace{-1em}
  \label{tab:analyzer-bugs}
  \begin{tabular}{clll}
    \toprule
    \textbf{Bug} & \textbf{System} & \textbf{Type} & \textbf{Status} \\
    \midrule

    \bug{25} & \textsc{Forge}   & crash       & confirmed \\
    \bug{26} & \textsc{hevm}    & crash       & fixed (\fix{P}) \\
    \bug{27} & \textsc{hevm}    & imprecision & fixed (\fix{Q}) \\
    \bug{28} & \textsc{hevm}    & crash       & fixed (\fix{R}) \\
    \bug{29} & \textsc{hevm}    & imprecision & fixed (\fix{S}) \\
    \bottomrule
  \end{tabular}
\end{wraptable}
During our experiments, we also observed 2 additional
discrepancies resembling miscompilations that had already been
fixed. One caused \solc to emit library placeholders for internal
libraries, similar to bug~\bug{11}; the other caused \solx to emit
opcodes unsupported by the selected EVM version. We exclude both from
our bug counts.

\tabref{tab:analyzer-bugs} summarizes bugs that \tool exposed in the
analyzers used during our evaluation.
Analyzer crashes are observed directly, while imprecision bugs
correspond to reported assertion violations that do not reproduce
under concrete execution.
We found 5 such bugs: 3 crashes and 2 imprecision
bugs, of which 4 have already been fixed. Overall, \tool
exposed 29 bugs across compilers and analyzers.

\textbf{RQ2: Fixed miscompilations.}
The four \solx miscompilations found by \tool (bugs~\bug{1}--\bug{4}
in \tabref{tab:bugs}) have all since been fixed. They expose two
recurring sources of semantic errors: three arise from incorrect
management of the EVM stack during code generation or optimization,
whereas the fourth results from a mismatch between Solidity arithmetic
semantics and LLVM IR semantics.

\begin{figure}[t]
  \centering

  \begin{subfigure}[b]{0.48\columnwidth}
    \begin{lstlisting}[language=Solidity]
assert(((new E()).isTrue()
  ? (new D()).h
  : (b ? (new D()).g : (new D()).g)).selector
  == D.g.selector);
    \end{lstlisting}
    \vspace{-0.5em}    
    \caption{Bug~\bug{1}: Branch folding.}
    \label{fig:solx-branch-folding}
  \end{subfigure}
  \hfill
  \begin{subfigure}[b]{0.48\columnwidth}
    \begin{lstlisting}[language=Solidity]
function f(bool, bool, bool, bool, bool,
           bool, bool, bool, bool, bool,
           bool, bool, bool, bool, bool)
  public pure returns (bool)
{
  return true;
}
    \end{lstlisting}
    \vspace{-0.5em}
    \caption{Bug~\bug{2}: Deep-stack operations.}
    \label{fig:solx-deep-stack}
  \end{subfigure}

  \vspace{0.5em}
  
  \begin{subfigure}[b]{0.48\columnwidth}
    \begin{lstlisting}[language=Solidity]
unchecked {
  int256 a = ((y / x) % (x % int256(3)));
  int256 b = ((y / x) % (x % int256(3)));
  assert(a == b);
}
    \end{lstlisting}
    \vspace{-0.5em}    
    \caption{Bug~\bug{3}: Signed remainder.}
    \label{fig:solx-signed-remainder}
  \end{subfigure}
  \hfill
  \begin{subfigure}[b]{0.48\columnwidth}
    \begin{lstlisting}[language=Solidity]
bool x = ((a || b) || ((e && c) || (d || e)))
         || (c || (e && (a && e)));
bool y = g(a, b, c, d, e)
         || ((e || c) && ((a && e) || c));
assert(x == y);
    \end{lstlisting}
    \vspace{-0.5em}    
    \caption{Bug~\bug{4}: Branch elimination.}
    \label{fig:solx-branch-elimination}
  \end{subfigure}
  \vspace{-0.5em}
  \caption{Reduced excerpts from the four \solx miscompilations found
    by \tool.}
  \vspace{-1.5em}
  \label{fig:solx-miscompilations}
\end{figure}

% dear-fowl
\emph{Bug~\bug{1}: Branch folding.}
The example in \figref{fig:solx-branch-folding} returns the selector of
a function chosen by nested conditional expressions. At the source
level, the assertion is straightforward: \code{E.isTrue()} always
returns false, and both branches of the inner conditional select
\code{D.g}. The expression must therefore evaluate to
\code{D.g.selector}, regardless of the value of \code{b}.

The miscompilation occurs in LLVM's \code{BranchFolder} pass, a
backend optimization that simplifies branches in the machine-code
control-flow graph. During this simplification, \solx failed to
preserve the correct EVM stack layout. The resulting bytecode therefore
violates an assertion that is always valid at the source level.

% mint-fox
\emph{Bug~\bug{2}: Deep-stack operations.}
The second miscompilation is triggered by the simple function in
\figref{fig:solx-deep-stack}, which takes 15 arguments and always
returns \code{true}. The EVM is stack based, and its \code{DUP} and
\code{SWAP} instructions provide only limited access to deeper stack
values. When compilation requires an access that cannot be expressed
directly, the \solc frontend represents it using pseudo-instructions
that must later be lowered to valid EVM bytecode.

In the non-via-IR pipeline, which compiles Solidity directly without
lowering through Yul, \solx handled one of these deep-stack operations
incorrectly and emitted bytecode containing an invalid opcode.
Executing the compiled function therefore fails instead of returning
\code{true}. The bug demonstrates how incorrect stack lowering in the
backend can turn a source program with unambiguous semantics into
invalid EVM bytecode.

% wired-sunfish
\emph{Bug~\bug{3}: Signed remainder.}
The example in \figref{fig:solx-signed-remainder} computes the same
arithmetic expression twice, so the assertion must hold for every
input. For \code{x = -1} and \code{y = INT_MIN}, however, the
computation reaches the signed-remainder corner case \code{INT_MIN \%
  -1}. Solidity and the EVM define the result of this operation, but
LLVM's \code{srem} operation has undefined behavior. \solx lowered the
Solidity operation directly to \code{srem} without handling this
semantic difference. The generated bytecode could therefore produce
different results for the two equivalent source computations, causing
the assertion to fail.

% stirred-martin
\emph{Bug~\bug{4}: Branch elimination.}
The fourth miscompilation arises when compiling the Boolean
computations in \figref{fig:solx-branch-elimination} through the
via-IR pipeline, which lowers Solidity through Yul before EVM code
generation. Although syntactically quite different, both computations
simplify to \code{a || b || c || d || e} and must therefore produce
the same result. During optimization, however, \solx removes a branch
without removing its corresponding condition from the EVM stack. The
stale value changes the stack layout seen by subsequent instructions,
causing the two computations to produce different Boolean results and
violate the assertion.

Together, these bugs affect distinct stages of \solx, from Solidity
lowering to LLVM optimization and EVM stack management.

\textbf{RQ3: Time to bug.}
This experiment evaluates how efficiently \tool exposes known and
newly discovered miscompilations. To attribute a finding to a
specific bug, we require a fixed compiler version that lets us
check whether the observed discrepancy disappears. The 4
\solx miscompilations from RQ2 have available fixes and can
therefore be used for patch-based attribution. To obtain a broader
set of targets, we complement them with known \solc
miscompilations.

Restricting the official Solidity bug list~\cite{SolidityKnownBugs},
as of September 1st, 2026, to bugs affecting \solc versions from 0.8.0
onward, the version range supported by our current implementation,
leaves 18 bugs. We classify 16 as miscompilations,
excluding \texttt{DataLocationChangeInInternalOverride} and
\texttt{UserDefinedValueTypesBug}. We use these 16 known \solc
miscompilations together with the 4 \solx miscompilations as
targets. Thus, the experiment targets 20 miscompilations in
total.

\begin{table*}[t]
  \centering
  \small
  \setlength{\tabcolsep}{3pt}
  \caption{Time-to-bug results. Min, median, and max TTB are computed
    over workers that exposed the bug within the 72-hour budget.  W
    reports successful workers (out of 10), and F the number of
    attributed failing programs.}
  \vspace{-1em}
  \label{tab:time-to-bug}
  \begin{tabularx}{\textwidth}{Xrrrrr}
    \toprule
    \textbf{Bug} & \textbf{Min TTB} & \textbf{Median TTB} &
    \textbf{Max TTB} & \textbf{W} & \textbf{F} \\
    \midrule

    \multicolumn{6}{l}{
      \emph{Previously known bugs in \solc}} \\
    \midrule

    \tcode{InlineAssemblyMemorySideEffects}
      & 4m 18s
      & 1d 10h 56m 15s
      & 2d 23h 59m 32s
      & 10/10 & 657 \\
    \tcode{KeccakCaching}
      & 3h 05m 41s
      & 1d 19h 25m 03s
      & 2d 19h 46m 20s
      & 9/10 & 18 \\
    \tcode{MissingSideEffectsOnSelectorAccess}
      & 7h 11m 47s
      & 1d 5h 04m 12s
      & 2d 19h 56m 04s
      & 3/10 & 3 \\
    \tcode{SignedImmutables}
      & 2d 8h 24m 45s
      & 2d 10h 31m 09s
      & 2d 12h 37m 32s
      & 2/10 & 2 \\
    \tcode{TransientStorageClearingHelperCollision}
      & 7h 33m 04s
      & 7h 33m 04s
      & 7h 33m 04s
      & 1/10 & 1 \\

    \midrule
    \multicolumn{6}{l}{
      \emph{Newly discovered bugs in \solx}} \\
    \midrule

    Bug~\bug{1}: Branch folding
      & 1h 13m 32s
      & 1d 10h 53m 58s
      & 2d 21h 20m 40s
      & 9/10 & 25 \\
    Bug~\bug{2}: Deep-stack operations
      & 7m 18s
      & 1d 10h 37m 34s
      & 2d 23h 55m 28s
      & 10/10 & 1309 \\
    Bug~\bug{3}: Signed remainder
      & 1d 19h 48m 51s
      & 1d 19h 48m 51s
      & 1d 19h 48m 51s
      & 1/10 & 1 \\
    Bug~\bug{4}: Branch elimination
      & 11h 25m 21s
      & 1d 7h 23m 51s
      & 2d 9h 40m 16s
      & 7/10 & 10 \\

    \bottomrule
  \end{tabularx}
\end{table*}

For every flagged program, we recompile it with the fixed compiler
version and run all applicable analyzers. We attribute a finding to
the target bug if at least one analyzer successfully checks the fixed
version and none exposes the assertion violation. If all analyzers
time out or cannot process the program, the result is inconclusive
and we do not attribute the finding to that bug.

For each vulnerable compiler version, we run 10 independent
workers with different random seeds for 72 hours, each using
one CPU core and 8 GB of memory. \tabref{tab:time-to-bug} reports the
bugs that \tool exposed within this budget. The first group contains
previously known \solc bugs, while the second contains the newly
discovered \solx bugs from \tabref{tab:bugs}. For each bug, the
minimum, median, and maximum TTB are computed over the workers that
expose the bug. W reports the number of successful workers, and F the
total number of attributed failing programs across all workers.
Target bugs not exposed within the budget are omitted.

Overall, \tool exposed 9 of the 20 target
miscompilations: 5 of the 16 previously known \solc bugs
and all 4 newly discovered \solx bugs. Among successful
workers, the median TTB ranges from 7h 33m 04s to 2d 10h
  31m 09s. For 6 of the 9 exposed bugs, the median TTB
is below 36 hours, and for 8 it is below 48
hours.

The 11 known \solc miscompilations not exposed within the
budget generally have highly specialized trigger conditions. These
include unusual storage layouts, complex recursive or inheritance
structures, carefully structured or malformed ABI data, and Yul
features that do not arise when \solc compiles Solidity source. Such
conditions are not systematically exercised by \tool.

\begin{table*}[t]
  \centering
  \small
  \setlength{\tabcolsep}{2.5pt}
  \caption{Ablation results for smart-contract transformations (SCT)
    and symbolic EMI (SEMI). Median TTB is computed over successful
    workers, and W reports successful workers (out of 10).
    $\Delta M$ reports the percentage change in median TTB and
    $\Delta W$ the change in successful workers relative to
    \tabref{tab:time-to-bug}. \solc bug names are abbreviated.
    Green indicates improvement, red degradation, and a dash no
    exposure within 72 hours.}
  \vspace{-1em}
  \label{tab:ablation}
  \begin{tabularx}{\textwidth}{
    X
    rrrr
    rrrr
    rrrr
  }
  \toprule
  \textbf{Bug}
  & \multicolumn{4}{c}{\textbf{SCT only}}
  & \multicolumn{4}{c}{\textbf{SEMI only}}
  & \multicolumn{4}{c}{\textbf{Neither}} \\
  \cmidrule(lr){2-5}
  \cmidrule(lr){6-9}
  \cmidrule(lr){10-13}

  & \textbf{Med. TTB} & $\boldsymbol{\Delta M}$
    & \textbf{W} & $\boldsymbol{\Delta W}$
  & \textbf{Med. TTB} & $\boldsymbol{\Delta M}$
    & \textbf{W} & $\boldsymbol{\Delta W}$
  & \textbf{Med. TTB} & $\boldsymbol{\Delta M}$
    & \textbf{W} & $\boldsymbol{\Delta W}$ \\
  \midrule

  \multicolumn{13}{l}{
    \emph{Previously known bugs in \solc}} \\
  \midrule

  \tcode{Inlin}
    & -- & -- & -- & \negative{-10}
    & 1d 12h 29m 41s & \negative{+4\%} & 10 & 0
    & -- & -- & -- & \negative{-10} \\

  \tcode{Kecca}
    & -- & -- & -- & \negative{-9}
    & 1d 17h 30m 20s & \positive{-4\%} & 9 & 0
    & -- & -- & -- & \negative{-9} \\

  \tcode{Missi}
    & 2d 2h 29m 48s & \negative{+74\%}
      & 4 & \positive{+1}
    & -- & -- & -- & \negative{-3}
    & -- & -- & -- & \negative{-3} \\

  \tcode{Signe}
    & -- & -- & -- & \negative{-2}
    & -- & -- & -- & \negative{-2}
    & -- & -- & -- & \negative{-2} \\

  \tcode{Trans}
    & 18h 09m 38s & \negative{+140\%}
      & 4 & \positive{+3}
    & -- & -- & -- & \negative{-1}
    & -- & -- & -- & \negative{-1} \\

  \midrule
  \multicolumn{13}{l}{
    \emph{Newly discovered bugs in \solx}} \\
  \midrule

  Bug~\bug{1}
    & 1d 7h 43m 29s & \positive{-9\%}
      & 10 & \positive{+1}
    & -- & -- & -- & \negative{-9}
    & 11h 19m 12s & \positive{-68\%}
      & 1 & \negative{-8} \\

  Bug~\bug{2}
    & 1d 11h 51m 15s & \negative{+4\%}
      & 10 & 0
    & 1d 17h 51m 50s & \negative{+21\%}
      & 10 & 0
    & 1d 10h 45m 05s & 0\%
      & 10 & 0 \\

  Bug~\bug{3}
    & -- & -- & -- & \negative{-1}
    & -- & -- & -- & \negative{-1}
    & -- & -- & -- & \negative{-1} \\

  Bug~\bug{4}
    & 1d 21h 45m 07s & \negative{+46\%}
      & 7 & 0
    & 1d 8h 44m 30s & \negative{+4\%}
      & 6 & \negative{-1}
    & 1d 15h 19m 38s & \negative{+25\%}
      & 6 & \negative{-1} \\

  \midrule
  \textbf{Bugs}
    & \multicolumn{4}{c}{\textbf{5}}
    & \multicolumn{4}{c}{\textbf{4}}
    & \multicolumn{4}{c}{\textbf{3}} \\
  \bottomrule
  \end{tabularx}
\end{table*}

\textbf{RQ4: Ablation.}  To evaluate the contribution of the two
transformation families, we repeat the time-to-bug experiments from
RQ3 with three ablated configurations. `SCT only' disables symbolic
EMI, `SEMI only' disables the smart-contract transformations, and
`Neither' disables both transformation families while retaining the
\circil-derived equivalence check. Thus, `Neither' removes the
transformations, not the oracle itself. We use the same 10
workers and 72-hour budget as in RQ3. \tabref{tab:ablation}
reports the ablated results and compares them with the configuration
using both transformation families in \tabref{tab:time-to-bug}. For
each bug, $\Delta M$ reports the percentage change in median TTB and
$\Delta W$ the absolute change in the number of successful
workers. Median TTB is computed over successful workers, so we
interpret it together with the number of workers exposing each bug.

Both transformation families contribute to bug coverage. While the
configuration using both exposes 9 bugs, SCT only exposes
5, SEMI only exposes 4, and Neither exposes
3. The two single-family configurations expose largely
different subsets: only bugs~\bug{2} and \bug{4} are exposed by
both. Moreover, \code{SignedImmutables} and bug~\bug{3} are not
exposed by any ablated configuration within the time budget. Thus,
neither transformation family subsumes the other, and enabling both
substantially increases the set of exposed bugs.

For bugs exposed by both the full and ablated configurations, the
full configuration generally also has a lower median TTB. Compared
with SCT only, it has a lower median TTB for four of the five shared
bugs; compared with SEMI only, for three of the four. With neither
transformation family, bug~\bug{4} has a higher median TTB and
bug~\bug{2} is essentially unchanged. Bug~\bug{1} is the exception:
its median TTB is 68\% lower, but only one worker exposes it,
compared with 9 in the full configuration. These results show
that the transformations substantially increase bug coverage while
generally maintaining or improving median TTB for bugs that remain
exposed.

% -----------------------------------------------------------
\subsection{Threats to Validity}
\label{sec:threats}
% -----------------------------------------------------------

Our experiments are randomized, so time to bug can vary across runs.
We mitigate this by using 10 independent workers with different
random seeds for each configuration and by reporting how many workers
expose each bug.

Our results depend on the supported compiler versions, configurations,
analyzers, and generated program features. Bugs requiring unsupported
or particularly rare feature combinations may therefore not be exposed
within the time budget. For RQ3 and RQ4, bug attribution additionally
depends on the availability of a fixed compiler version; we use the
fix to check that the observed discrepancy disappears.

%%% Local Variables:
%%% mode: latex
%%% TeX-master: "main"
%%% End:

%% file: related.tex
%!TEX root = main.tex

% -----------------------------------------------------------
\section{Related Work}
\label{sec:related}
% -----------------------------------------------------------

\textbf{Compiler testing and semantic oracles.}
Compiler testing has a long history of generating programs that
exercise compiler behavior and checking the resulting executables for
discrepancies; Chen et al.~\cite{ChenPatra2020} provide a
comprehensive survey of this area. Random program generators such as
\textsc{Csmith}~\cite{YangChen2011} generate programs with
well-defined semantics and use differential
testing~\cite{McKeeman1998} across compilers to expose
miscompilations. \textsc{YARPGen}~\cite{LivinskiiBabokin2020} targets
compiler optimizers by guiding generation toward constructs likely to
trigger optimization passes. Skeletal program
generation~\cite{ZhangSun2017} instead derives program families from
existing compiler tests by systematically filling holes with in-scope
variables. These approaches rely on differential testing against
another compiler, version, or configuration, and can therefore miss
bugs shared by both sides of the comparison.

Metamorphic testing~\cite{ChenCheung1998} instead derives relations
between executions that can serve as semantic oracles without
requiring a known expected output. Equivalence Modulo Inputs (EMI)
applies this idea to compiler testing. Its original realization,
\textsc{Orion}~\cite{LeAfshari2014}, removes code that is dead for a
particular input, while \textsc{Athena}~\cite{LeSun2015-Compilers} and
\textsc{Hermes}~\cite{SunLe2016-Mutation} extend EMI with additional
semantics-preserving transformations. These techniques derive
equivalence from coverage observed for concrete executions. In
contrast, symbolic EMI embeds the coverage relation in the generated
program, yielding a single equivalence oracle over the input space
explored by the analyzer.

\textbf{Testing smart-contract compilers.}
Ma et al.~\cite{MaZhang2024} studied bugs in \solc and found that a
substantial fraction involved incorrect output, highlighting the need
for semantic oracles that can detect miscompilations.
\acf~\cite{GroceVanTonder2022} applies compiler-oriented mutation and
splicing, while \fuzzol~\cite{MitropoulosSotiropoulos2023} introduces
syntax-aware mutations over Solidity's AST and prioritizes mutation
strategies that increase compiler coverage. Both primarily rely on
compiler failures as bug oracles and therefore cannot reliably detect
miscompilations. \erwin~\cite{MaDonaldson2026} instead generates
Solidity program templates and systematically enumerates values for
their placeholders, but similarly targets robustness failures and
invalid-program acceptance rather than semantic discrepancies in
successfully compiled bytecode.

Li et al.'s \solsmith~\cite{LiLiu2025} and Shastry's
\efsolsmith~\cite{Shastry2026} target miscompilations by comparing
generated Solidity programs across compiler configurations.
\textsc{DeSCDT}~\cite{TianWang2024} combines deep-learning-based
generation and mutation with optimized-versus-unoptimized comparison.
\textsc{ParityFuzz}~\cite{SuYe2026} uses LLM-generated mutation rules
to produce test programs and compares their behavior across different
Solidity compilers. \textsc{IDOL}~\cite{LiLiang2025} combines
cross-optimization comparison with reverse-optimization
transformations that generate equivalent Solidity variants. All these
approaches rely on differential oracles and can therefore miss bugs
that affect both sides of the comparison identically. In contrast,
\tool embeds equivalence checks directly in generated programs and
requires no reference compiler configuration. Our goal is therefore
not to replace differential testing, but to provide a complementary
oracle that does not require a reference compiler, pipeline, or
configuration.

\textsc{sBugChecker}~\cite{TongLi2025} and
\textsc{CompileGuard}~\cite{YeChen2026} address a different problem:
they analyze contracts for conditions that trigger known compiler bugs
rather than testing compilers for new bugs.

\textbf{Metamorphic testing of specialized compilation pipelines.}
\textsc{Circuzz}~\cite{HochrainerIsychev2025} applies metamorphic
testing to zero-knowledge circuit toolchains. It uses \circil to
generate semantically equivalent expressions, lowers them through the
target toolchain, and compares their results on common inputs.
\textsc{Arguzz}~\cite{HochrainerWuestholz2026} targets zkVMs using
self-checking product programs that compare semantically equivalent
computations. \tool instead uses self-checking programs as targets for
analyzer-driven exploration: after compilation, existing fuzzers and
symbolic execution engines search the resulting EVM bytecode for
inputs that violate the embedded assertions.

\textbf{Testing program analyzers.}
Program analyzers have themselves been tested using differential,
specification-based, and metamorphic
oracles~\cite{KlingerChristakis2019,KapusCadar2017,BugariuWuestholz2018,TanejaLiu2020,MordahlZhang2023,EvenMendozaSharma2023,ZhangPei2023,HeDi2024,KaindlstorferIsychev2024,FleischmannKaindlstorfer2024}. These
techniques expose soundness, precision, and implementation bugs in
static analyzers and symbolic execution engines. Although \tool uses
analyzers primarily to explore the generated equivalence checks, this
exploration can also expose bugs in the analyzers themselves.

%%% Local Variables:
%%% mode: latex
%%% TeX-master: "main"
%%% End:

%% file: conclusion.tex
%!TEX root = main.tex

% -----------------------------------------------------------
\section{Conclusion}
\label{sec:conclusion}
% -----------------------------------------------------------

We presented \tool, an approach for finding miscompilations in EVM
smart-contract compilers using assertion-driven equivalence oracles.
\tool combines smart-contract transformations with symbolic EMI and
lets existing analyzers search compiled bytecode for assertion
violations. Across three compilers, \tool found 24 bugs,
including 6 miscompilations, and our ablations show that the
two transformation families are complementary, substantially
increasing bug coverage while generally maintaining or improving
median TTB.

%%% Local Variables:
%%% mode: latex
%%% TeX-master: "main"
%%% End:

%% file: bibliography.bib
@inproceedings{BartheCrespo2011,
  author    = {Gilles Barthe and Juan Manuel Crespo and C{\'{e}}sar
                  Kunz},
  title     = {Relational Verification Using Product Programs},
  booktitle = {FM},
  series    = {LNCS},
  volume    = {6664},
  year      = {2011},
  pages     = {200--214},
  publisher = {Springer},
  COMMENTurl       = {https://doi.org/10.1007/978-3-642-21437-0\_17},
  COMMENTdoi       = {10.1007/978-3-642-21437-0\_17}
}

@inproceedings{BartheDArgenio2004,
  author    = {Gilles Barthe and Pedro R. D'Argenio and Tamara Rezk},
  title     = {Secure Information Flow by Self-Composition},
  booktitle = {CSFW},
  year      = {2004},
  pages     = {100--114},
  publisher = {IEEE Computer Society}
}

@inproceedings{BugariuWuestholz2018,
  author    = {Bugariu, Alexandra and W\"ustholz, Valentin and
                  Christakis, Maria and M\"uller, Peter},
  title     = {Automatically Testing Implementations of Numerical
                  Abstract Domains},
  booktitle = {ASE},
  year      = {2018},
  pages     = {768--778},
  publisher = {ACM}
}

@techreport{ChenCheung1998,
  author      = {Chen, Tsong Yueh and Cheung, S. C. and Yiu,
                  Siu{-}Ming},
  title       = {Metamorphic Testing: {A} New Approach for Generating
                  Next Test Cases},
  year        = {1998},
  number      = {HKUST--CS98--01},
  institution = {HKUST}
}

@article{ChenPatra2020,
  author    = {Chen, Junjie and Patra, Jibesh and Pradel, Michael and
                  Xiong, Yingfei and Zhang, Hongyu and Hao, Dan and
                  Zhang, Lu},
  title     = {A Survey of Compiler Testing},
  journal   = {Comput. Surv.},
  volume    = {53},
  issue     = {1},
  year      = {2020},
  pages     = {4:1--4:36},
  publisher = {ACM}
}

@inproceedings{ClarksonSchneider2008,
  author    = {Clarkson, Michael R.  and Schneider, Fred B.},
  title     = {Hyperproperties},
  booktitle = {CSF},
  year      = {2008},
  pages     = {51--65},
  publisher = {IEEE Computer Society}
}

@inproceedings{DxoSoos2024,
  author    = {Dxo and Mate Soos and Zoe Paraskevopoulou and Martin
                  Lundfall and Mikael Brockman},
  title     = {{Hevm}, a Fast Symbolic Execution Framework for {EVM} Bytecode},
  booktitle = {CAV},
  series    = {LNCS},
  volume    = {14681},
  year      = {2024},
  pages     = {453--465},
  publisher = {Springer}
}

@misc{Echidna,
  title     = {Echidna},
  note      = {\url{https://github.com/crytic/echidna}}
}

@inproceedings{EvenMendozaSharma2023,
  author    = {Karine Even{-}Mendoza and Arindam Sharma and Alastair
                  F. Donaldson and Cristian Cadar},
  title     = {{GrayC}: {G}reybox Fuzzing of Compilers and Analysers
                  for {C}},
  booktitle = {ISSTA},
  year      = {2023},
  pages     = {1219--1231},
  publisher = {ACM}
}

@inproceedings{FleischmannKaindlstorfer2024,
  author    = {Fleischmann, Markus and Kaindlstorfer, David and
                  Isychev, Anastasia and W\"{u}stholz, Valentin and
                  Christakis, Maria},
  title     = {Constraint-Based Test Oracles for Program Analyzers},
  booktitle = {ASE},
  year      = {2024},
  pages     = {344-355},
  publisher = {ACM}
}

@misc{Foundry,
  title     = {Foundry},
  note      = {\url{https://github.com/foundry-rs/foundry}}
}

@inproceedings{GriecoSong2020,
  author    = {Grieco, Gustavo and Song, Will and Cygan, Artur and
                  Feist, Josselin and Groce, Alex},
  title     = {Echidna: {E}ffective, Usable, and Fast Fuzzing for
                  Smart Contracts},
  booktitle = {ISSTA},
  year      = {2020},
  pages     = {557--560},
  publisher = {ACM}
}

@inproceedings{GroceVanTonder2022,
  author    = {Alex Groce and Rijnard van Tonder and Goutamkumar
                  Tulajappa Kalburgi and Claire {Le Goues}},
  title     = {Making No-Fuss Compiler Fuzzing Effective},
  booktitle = {CC},
  year      = {2022},
  pages     = {194--204},
  publisher = {ACM}
}

@article{HeDi2024,
  author    = {Weigang He and Peng Di and Mengli Ming and Chengyu
                  Zhang and Ting Su and Shijie Li and Yulei Sui},
  title     = {Finding and Understanding Defects in Static Analyzers by Constructing
                  Automated Oracles},
  journal   = {PACMSE},
  volume    = {1},
  issue     = {FSE},
  year      = {2024},
  pages     = {1656--1678},
  publisher = {ACM}
}

@inproceedings{HochrainerIsychev2025,
  author    = {Christoph Hochrainer and Anastasia Isychev and Valentin
                  W\"{u}stholz and Maria Christakis},
  title     = {Fuzzing Processing Pipelines for Zero-Knowledge
                  Circuits},
  booktitle = {CCS},
  year      = {2025},
  pages     = {783--797},
  publisher = {ACM}
}

@inproceedings{HochrainerWuestholz2026,
  author    = {Christoph Hochrainer and Valentin W{\"{u}}stholz and
                  Maria Christakis},
  title     = {{Arguzz}: {T}esting {zkVMs} for Soundness and
                  Completeness Bugs},
  booktitle = {Security},
  year      = {2026},
  pages     = {1807--1826},
  publisher = {USENIX}
}

@inproceedings{KaindlstorferIsychev2024,
  author    = {Kaindlstorfer, David and Isychev, Anastasia and
                  W\"{u}stholz, Valentin and Christakis, Maria},
  title     = {Interrogation Testing of Program Analyzers for
                  Soundness and Precision Issues},
  year      = {2024},
  booktitle = {ASE},
  pages     = {319-330},
  publisher = {ACM}
}

@inproceedings{KapusCadar2017,
  author    = {Kapus, Timotej and Cadar, Cristian},
  title     = {Automatic Testing of Symbolic Execution Engines via
                  Program Generation and Differential Testing},
  booktitle = {ASE},
  year      = {2017},
  pages     = {590--600},
  publisher = {IEEE Computer Society}
}

@inproceedings{KlingerChristakis2019,
  author    = {Klinger, Christian and Christakis, Maria and
                  W{\"{u}}stholz, Valentin},
  title     = {Differentially Testing Soundness and Precision of
                  Program Analyzers},
  booktitle = {ISSTA},
  year      = {2019},
  pages     = {239--250},
  publisher = {ACM}
}

@inproceedings{LeAfshari2014,
  author    = {Le, Vu and Afshari, Mehrdad and Su, Zhendong},
  title     = {Compiler Validation via Equivalence Modulo Inputs},
  booktitle = {PLDI},
  year      = {2014},
  pages     = {216--226},
  publisher = {ACM}
}

@inproceedings{LeSun2015-Compilers,
  author    = {Le, Vu and Sun, Chengnian and Su, Zhendong},
  title     = {Finding Deep Compiler Bugs via Guided Stochastic Program Mutation},
  booktitle = {OOPSLA},
  year      = {2015},
  pages     = {386--399},
  publisher = {ACM}
}

@inproceedings{LiLiang2025,
  author    = {Lantian Li and Yejian Liang and Zhongxing Yu},
  title     = {{IDOL}: {I}mproved Different Optimization Levels
                  Testing for {Solidity} Compilers},
  booktitle = {QRS Companion},
  year      = {2025},
  pages     = {773--774},
  publisher = {IEEE Computer Society}
}

@inproceedings{LiLiu2025,
  author    = {Lantian Li and Zhihao Liu and Zhongxing Yu},
  title     = {{Solsmith}: {Solidity} Random Program Generator for
                  Compiler Testing},
  booktitle = {QRS Companion},
  year      = {2025},
  pages     = {178--187},
  publisher = {IEEE Computer Society}
}

@article{LivinskiiBabokin2020,
  author    = {Livinskii, Vsevolod and Babokin, Dmitry and Regehr,
                  John},
  title     = {Random Testing for {C} and {C++} Compilers with
                  {YARPGen}},
  journal   = {PACMPL},
  volume    = {4},
  issue     = {OOPSLA},
  year      = {2020},
  pages     = {196:1--196:25},
  publisher = {ACM}
}

@inproceedings{MaDonaldson2026,
  author    = {Haoyang Ma and Alastair F. Donaldson and Qingchao Shen
                  and Yongqiang Tian and Junjie Chen and Shing-Chi
                  Cheung},
  title     = {Bounded Exhaustive Random Program Generation for
                  Testing {Solidity} Compilers},
  booktitle = {ICSE},
  year      = {2026},
  pages     = {3410--3422},
  publisher = {ACM}
}

@inproceedings{MaZhang2024,
  author    = {Haoyang Ma and Wuqi Zhang and Qingchao Shen and
                  Yongqiang Tian and Junjie Chen and Shing{-}Chi
                  Cheung},
  title     = {Towards Understanding the Bugs in {Solidity} Compiler},
  booktitle = {ISSTA},
  year      = {2024},
  pages     = {1312--1324},
  publisher = {ACM}
}

@article{McKeeman1998,
  author    = {McKeeman, William M.},
  title     = {Differential Testing for Software},
  journal   = {Digital Technical Journal},
  volume    = {10},
  issue     = {1},
  year      = {1998},
  pages     = {100--107},
  publisher = {ACM}
}

@misc{Medusa,
  title     = {Medusa},
  note      = {\url{https://github.com/crytic/medusa}}
}

@inproceedings{MitropoulosSotiropoulos2023,
  author    = {Charalambos Mitropoulos and Thodoris Sotiropoulos and
                  Sotiris Ioannidis and Dimitris Mitropoulos},
  title     = {Syntax-Aware Mutation for Testing the {Solidity} Compiler},
  booktitle = {ESORICS},
  series    = {LNCS},
  volume    = {14346},
  year      = {2023},
  pages     = {327--347},
  publisher = {Springer}
}

@inproceedings{MordahlZhang2023,
  author    = {Austin Mordahl and Zenong Zhang and Dakota Soles and
                  Shiyi Wei},
  title     = {{ECSTATIC}: {A}n Extensible Framework for Testing and
                  Debugging Configurable Static Analysis},
  booktitle = {ICSE},
  year      = {2023},
  pages     = {550--562},
  publisher = {IEEE Computer Society}
}

@article{Shastry2026,
  author    = {Bhargava Shastry},
  title     = {Finding and Understanding Miscompilation Bugs in the
                  {Solidity} Compiler},
  journal   = {CoRR},
  volume    = {abs/2607.07217},
  year      = {2026}
}

@misc{Solidity,
  title     = {Solidity},
  note      = {\url{https://github.com/ethereum/solidity}}
}

@misc{SolidityKnownBugs,
  title     = {{Solidity} Compiler: {L}ist of Known Bugs},
  note      = {\url{https://docs.soliditylang.org/en/latest/bugs.html}}
}

@inproceedings{SunLe2016-Mutation,
  author    = {Sun, Chengnian and Le, Vu and Su, Zhendong},
  title     = {Finding Compiler Bugs via Live Code Mutation},
  booktitle = {OOPSLA},
  year      = {2016},
  pages     = {849--863},
  publisher = {ACM}
}

@article{SuYe2026,
  author    = {Bowei Su and Mingxi Ye and Yuhong Nan and Peilin Zheng
                  and Zibin Zheng},
  title     = {{ParityFuzz}: {F}inding Inconsistencies across
                  {Solidity} Compilers via Fine-Grained Mutation and
                  Differential Analysis},
  journal   = {J. Syst. Softw.},
  volume    = {241},
  issue     = {C},
  year      = {2026},
  pages     = {112927},
  publisher = {Elsevier}
}

@inproceedings{TanejaLiu2020,
  author    = {Taneja, Jubi and Liu, Zhengyang and Regehr, John},
  title     = {Testing Static Analyses for Precision and Soundness},
  booktitle = {CGO},
  year      = {2020},
  pages     = {81--93},
  publisher = {ACM}
}

@inproceedings{TerauchiAiken2005,
  author    = {Tachio Terauchi and Alex Aiken},
  title     = {Secure Information Flow as a Safety Problem},
  booktitle = {SAS},
  series    = {LNCS},
  volume    = {3672},
  year      = {2005},
  pages     = {352--367},
  publisher = {Springer},
  COMMENTurl       = {https://doi.org/10.1007/11547662\_24},
  COMMENTdoi       = {10.1007/11547662\_24}
}

@article{TianWang2024,
  author    = {Zhenzhou Tian and Fanfan Wang and Yanping Chen and
                  Lingwei Chen},
  title     = {Differential Testing {Solidity} Compiler Through Deep
                  Contract Manipulation and Mutation},
  journal   = {Softw. Qual. J.},
  volume    = {32},
  issue     = {2},
  year      = {2024},
  pages     = {765--790},
  publisher = {Springer}
}

@article{TongLi2025,
  author    = {Fei Tong and Zihao Li and Guang Cheng and Yujian Zhang
                  and Heng Li},
  title     = {{sBugChecker}: {A} Systematic Framework for Detecting
                  {Solidity} Compiler-Introduced Bugs},
  journal   = {Trans. Inf. Forensics Secur.},
  volume    = {20},
  year      = {2025},
  pages     = {9760--9775},
  publisher = {IEEE Computer Society}
}

@inproceedings{YangChen2011,
  author    = {Yang, Xuejun and Chen, Yang and Eide, Eric and Regehr,
                  John},
  title     = {Finding and Understanding Bugs in {C} Compilers},
  booktitle = {PLDI},
  year      = {2011},
  pages     = {283--294},
  publisher = {ACM}
}

@article{YeChen2026,
  author    = {Zhifan Ye and Jiachi Chen and Zhongxin Liu and Chong
                  Chen and Xiapu Luo and Ting Chen and Difan Xie and
                  Xiaohu Yang},
  title     = {When Source Code Misleads: {D}etecting Compiler Version
                  Risks in Smart Contracts},
  journal   = {Blockchain: Research and Applications},
  year      = {2026},
  pages     = {100548},
  publisher = {Elsevier}
}

@inproceedings{ZhangPei2023,
  author    = {Huaien Zhang and Yu Pei and Junjie Chen and Shin Hwei
                  Tan},
  title     = {Statfier: {A}utomated Testing of Static Analyzers via
                  Semantic-Preserving Program Transformations},
  booktitle = {ESEC/FSE},
  year      = {2023},
  pages     = {237--249},
  publisher = {ACM}
}

@inproceedings{ZhangSun2017,
  author    = {Zhang, Qirun and Sun, Chengnian and Su, Zhendong},
  title     = {Skeletal Program Enumeration for Rigorous Compiler
                  Testing},
  booktitle = {PLDI},
  year      = {2017},
  pages     = {347--361},
  publisher = {ACM}
}
